\documentclass[aps,prapplied,twocolumn,superscriptaddress]{revtex4-2}%
\usepackage{graphics}
\usepackage{subfigure}
\usepackage{amsmath}
\usepackage{physics}
\usepackage{graphicx}
\usepackage{color}
\usepackage{amsfonts}
\usepackage{amssymb}
\usepackage{float}
\usepackage{longtable}
\usepackage{epsfig}
\usepackage{latexsym}
\usepackage{theorem}
\usepackage{bbm}
\usepackage{hyperref}
\usepackage{comment}

\usepackage{bm}
\usepackage{psfrag}
\usepackage{bm}
\providecommand{\U}[1]{\protect\rule{.1in}{.1in}}

\definecolor{LBc}{RGB}{134,41,198}

\DeclareMathOperator*{\ave}{\mathbb{E}}
\DeclareMathOperator*{\argmax}{argmax}
\DeclareMathOperator*{\argmin}{argmin}

\newcommand{\bs}[1]{\boldsymbol{#1}}

\begin{document}

\title{Quantum-enhanced barcode decoding and pattern recognition }
\author{Leonardo Banchi}
\affiliation{Department of Physics and Astronomy, University of Florence,
via G. Sansone 1, I-50019 Sesto Fiorentino (FI), Italy}
\affiliation{
INFN Sezione di Firenze, via G. Sansone 1, I-50019, Sesto Fiorentino (FI), Italy
}
\author{Quntao Zhuang}
\affiliation{
Department of Electrical and Computer Engineering, University of Arizona, Tucson, Arizona 85721, USA
}
\affiliation{
James C. Wyant College of Optical Sciences, University of Arizona, Tucson, Arizona 85721, USA
}
\author{Stefano Pirandola}
\affiliation{Department of Computer Science, University of York, York YO10 5GH, UK}
\date{\today}

\begin{abstract}
Quantum hypothesis testing is one of the most fundamental problems in quantum information theory, with crucial implications in areas like quantum sensing, where it has been used to prove quantum advantage in a series of binary photonic protocols, e.g., for target detection or memory cell readout. In this work, we generalize this theoretical model to the multi-partite setting of barcode decoding and pattern recognition. We start by defining a digital image as an array or grid of pixels, each pixel corresponding to an ensemble of quantum channels. Specializing each pixel to a black and white alphabet, we naturally define an optical model of barcode. In this scenario, we show that the use of quantum entangled sources, combined with suitable measurements and data processing, greatly outperforms classical coherent-state strategies for the tasks of barcode data decoding and classification of black and white patterns. 
Moreover, introducing relevant bounds, we show that the problem of pattern recognition is significantly simpler than 
barcode decoding, as long as the minimum Hamming distance between images from different classes is 
	large enough. 
Finally, 
we theoretically demonstrate the advantage of using quantum sensors for 
pattern recognition with the nearest neighbor classifier, a supervised learning algorithm, and 
numerically verify this prediction for handwritten digit classification. 
\end{abstract}

\maketitle

\section{Introduction}
Quantum information science has undergone remarkable advances in the recent years, progressing in all its sub-fields of computing~\cite{Nielsen_Chuang,preskill2018quantum}, communication~\cite{ReviewQKD2020} and sensing~\cite{Cappellaro_2012,ReviewSensing}, both with discrete- and continuous-variable systems~\cite{BraunsteinREV,SerafiniBook,Weedbrook_2012}. In this wide scenario, quantum sensing is arguably one of the most mature areas for near-term technological deployment. Its theoretical and experimental developments have been strongly based on quantum metrology~\cite{MetrologySAM,Paris09} and quantum hypothesis testing~\cite{Helstrom_1976,Barnett_09,Chefles_2000,Janos2010}. Especially, the latter approach has allowed to show a quantum advantage over classical strategies in tasks of target detection~\cite{Lloyd2008,Tan2008} and data readout~\cite{Qreading}, modelled as binary problems of quantum channel discrimination. 

The discrimination of quantum channels is an incredibly rich area of investigation~\cite{zhuang2019physical,zhuang2020entanglement}, with unexplored consequences but also non-trivial difficulties. It represents a double optimization problem where both input states and output measurements need to be varied. Furthermore, in the bosonic setting, it has to be formulated as an energy-constrained problem, where the mean number of input photons is limited to some finite, small, value. In such a scenario, the central question is that of showing quantum advantage: Can truly-quantum states, e.g., entangled, lead to an advantage over classical, i.e. coherent, states? Addressing this question in the multi-ary case is difficult, since the theory is missing powerful tools that are instead available for the binary case. 


In this work, we take a step forward by developing the theory of quantum hypothesis testing for the multi-ary setting of barcode decoding and pattern classification. We start from a general model of digital image, where each pixel is described by an ensemble of quantum channels defined over a finite alphabet. We specialize to the case where the single-pixel alphabet is binary, so that there are $2^n$ possible hypotheses or configurations for an $n$-pixel barcode. We then show how the use of quantum sources of light, based on entangled states, can clearly outperform classical strategies based on coherent states for the readout of the barcode configuration, i.e., the retrieval of its data. In particular, we derive an analytical condition for the maximum number of pixels or the minimum number of probings such that quantum advantage is obtained. This result holds not only for a uniform distribution of the possible configurations, but also when data is stored by the positions of $k$ white pixels among a grid of otherwise black pixels.

Besides data readout or barcode decoding, we consider the general problem of pattern recognition, 
where the task is to classify  an image, e.g.~a handwritten digit, without necessarily reconstructing it 
pixel by pixel. Here the image distribution 
is not uniform and generally unknown, and optimal classification has to be approximated via a collection 
of correctly classified examples, following supervised learning strategies \cite{murphy2012machine}. 
We consider the ultimate limits of this procedure, where we may optimize over the optical circuit, measurements 
and subsequent classical post-processing algorithms. Introducing relevant bounds, we theoretically prove that this 
problem is significantly simpler than that of barcode decoding, as long as the minimum Hamming distance between 
images from different classes is large enough. Moreover, we show how clear quantum advantage can be 
obtained as a function of the number of training data. 
Finally, we consider a simplified scheme for recognizing black and white
patterns, such as digital images of handwritten digits, by means local
measurements followed by a classical nearest neighbor classifier. More
specifically, we apply this classifier to the measurement outcomes that are
obtained by either using entangled states or coherent states at the input of
the grid of pixels. We are able to show a clear quantum advantage that holds
even when we employ sub-optimal photon counting measurements for the quantum
case, which are particularly relevant for near-term experiments. 
The advantage becomes particularly evident at relatively
small energies where a total of a few hundred of photons are irradiated over
each pixel.

The paper is organized as follows. In Section~\ref{s:barcode} we discuss the problem of 
barcode decoding and show the possible advantage in using quantum detectors with 
entangled input states. In Section~\ref{s:pattern} we discuss the related problem 
of pattern recognition, showing a similar advantage. Discussions are drawn in Section~\ref{s:conclusions}.

\section{Barcode Decoding}\label{s:barcode}

\subsection{Quantum mechanical model of a digital image}

A basic imaging system irradiates light over an array of pixels which can be read in transmission or in reflection. From the ratios between input and output intensities, the system generates a corresponding array of grey-levels that constitutes a monochromatic image. In a quantum mechanical setting, each pixel can therefore be modeled as a bosonic lossy channel $\mathcal E_i$ whose transmissivity depends on the grey-level $i$. This lossy channel can be probed by an input state (with some limited energy) and a corresponding output measurement, generally described by positive-operator valued measure (POVM). Finally, the outcome is processed by a decision test that identifies the channel and, therefore, the grey-level $i$.

Let us formalize the problem in more mathematical detail, which can be seen as a multi-ary and multi-pixel generalization of the basic model of quantum reading~\cite{Qreading}. Assume that each pixel is described by a channel ensemble $\{\mathcal{E}%
_{i}\}$ spanned by the label $0\leq i\leq C-1$, where $C$ is the discrete number of grey-levels that can 
be assumed by the pixel. Let us define
an {\it image} over $n$\ pixels as a sequence $\bs{i}:=i_{0},\cdots,i_{n-1}$, together with an associated probability distribution $\pi_{\bs{i}}$, which is simply $\pi_{\bs{i}}=C^{-n}$ in the uniform case. 
The global channel describing the entire array of $n$ pixels is the tensor product 
$\mathcal{E}_{\bs{i}}^{n}:=\mathcal{E}_{i_{0}}\otimes\cdots\otimes
\mathcal{E}_{i_{n-1}}$. Thus, an image can equivalently be represented by an ensemble
of multi-channels $\{\pi_{\bs{i}},\mathcal{E}_{\bs{i}}^{n}\}$.


In order to read the image, let us assume that we have a generic $2n$-mode state $\tilde\rho$ at the input: $n$ signal modes are sent through the pixels, while $n$ idler modes are used to help the measurement. At the output, there is an ensemble of possible states $\{\pi_{\bs{i}}%
,\rho_{\bs{i}}\}$ where $\rho_{\bs{i}}:=\mathcal I^{n}\otimes \mathcal{E}_{\bs{i}}%
^{n}(\tilde\rho)$. In the case of a classical transmitter, the signal modes are prepared in coherent states while the idler modes are in vacuum states. In the case of a quantum transmitter, signal and idler modes are entangled pairwise. In particular, each signal-idler pair is described by a two-mode squeezed vacuum (TMSV) state.




In general, we probe the image $\bs{i}$ with identical inputs for $M$ times, leading to the overall input state $\tilde\rho^{\otimes M}$ and corresponding output 
$\rho_{\bs{i}}^{\otimes M}:=\left[\mathcal I^n\otimes  \mathcal{E}_{\bs{i}}^{n}%
(\tilde\rho)\right]  ^{\otimes M}$. We measure this output with a collective POVM,
where each measurement operator $\Pi_{\bs{i}'}$ represents the decision that the image is $\bs{i}'$. Because input states are energy-constrained, there will be readout errors described by the conditional probabilities
\begin{equation}
	p_{\rm read}({\bs i}'|\bs i) = \Tr\left(\Pi_{{\bs i}'}\rho_{\bs i}^{\otimes M}\right)~.
	\label{measupattern}
\end{equation}
By including the priors $\{\pi_i\}$, we may therefore define the success probability or, equivalently, the error probability
\begin{align}
	p_{\text{succ}}&:=\sum_{\bs{i}}\pi_{\bs{i}}p_{\rm read}(\bs{i}%
|\bs{i}),~~~p_{\text{err}}=1-p_{\text{succ}}.
\label{success}
\end{align}

Using Refs.~\cite{Barnum,Montanaro} and the multiplicativity of the fidelity over tensor products, one finds that the minimum error probability (optimized over POVMs) satisfies
\begin{equation}
\frac{1}{2}\sum_{\bs{i}\neq\bs{j}}\pi_{\bs{i}%
}\pi_{\bs{j}}F^{2M}_{{\bs{i}}:{\bs{j}}}
\leq
p_{\text{err}}\leq\sum_{\bs{i}\neq\bs{j}}\sqrt{\pi_{\bs{i}}%
\pi_{\bs{j}}}F^{M}_{{\bs{i}}:{\bs{j}}},
 \label{BarnumMontanaro}%
\end{equation}
where  
\begin{equation}
F_{{\bs{i}}:{\bs{j}}} := 
F(\rho_{\bs{i}},\rho_{\bs{j}})=
\Vert\sqrt{\rho_{\bs i}}\sqrt{\rho_{\bs j}}\Vert_{1}%
=\mathrm{Tr}\sqrt{\sqrt{\rho_{\bs i}}\rho_{\bs j}\sqrt{\rho_{\bs i}}}%
\end{equation}
is the fidelity between two generic single-probing multi-pixel output states, $\rho_{\bs{i}}$ and $\rho_{\bs{j}}$. 
The inequalities in Eq.~\eqref{BarnumMontanaro} bound the performances of a pretty good measurement~\cite{PGM1,PGM2,PGM3} and have no explicit dependence on the dimension of the Hilbert space, so that they hold for bosonic states as long as these states are energy-constrained. 
Below, we build on these inequalities to derive our bounds for decoding barcodes.






\subsection{Barcode discrimination}

\begin{figure}[th!]
	\centering
	\includegraphics[width=0.95\linewidth]{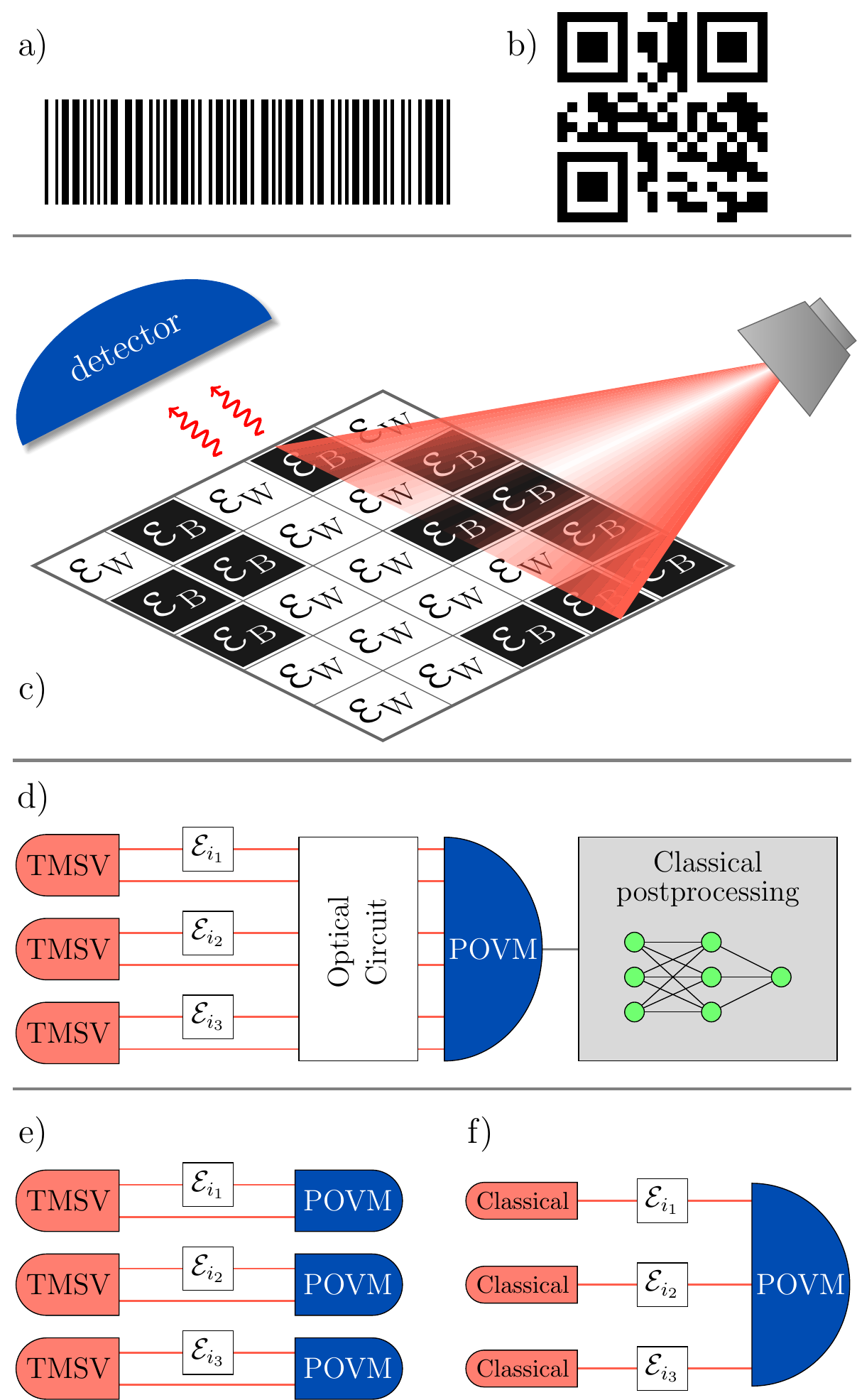}
	\caption{ { Barcode decoding.} Examples of 1D (a) and 2D (b) barcodes.  (c) 
		Schematic physical setup for decoding a 2D barcode with $n=5\times5$ pixels. A source shines light on the grid of pixels, each modeled by a quantum channel, which is either $\mathcal E_B$ or $\mathcal E_W$ depending on the 
		pixel grey-level, black (B) or white (W). The reflected/scattered light is collected by a detector, which aims at recognizing {\it all} pixel values depending on 
		the detected photons. In a quantum setup, the signal modes shined over the pixels are entangled with idler modes that are directly sent to the detector for a joint measurement.  
		(d) General theoretical setup scheme for optimal barcode discrimination using entangled TMSV states, 
		where the mixing optical circuit, the measurements and following classical post-processing must be optimized. 
		(e,f) Special setup to claim quantum advantage: we compare independent entangled TMSV states on each pixel followed 
		by independent {\it local} measurements (e) with classical coherent sources
		followed by global measurements. Quantum advantage is claimed whenever (e) beats (f).
	}
	\label{fig:barcode}
\end{figure}

An important case of the general problem discussed in the previous section is barcode decoding, whose 
schematic setup is shown in Fig.~\ref{fig:barcode}. 
A {\it barcode} is either a 
one-dimensional (1D) or two-dimensional (2D) grid 
of pixels with two possible colors, black (B) or white (W). 
With a slight abuse of jargon, we call {\it pixels} the elements 
of the 1D or 2D grid that define a barcode. In the 1D case (Fig.~\ref{fig:barcode}a),
a pixel is a black or white vertical bar, while in the 2D case (Fig.~\ref{fig:barcode}b)
a pixel is an elementary square. It is worth noting that many of the conclusions drawn for barcode decoding can be extended to more general images. Indeed, 
a higher number of grey-levels $C>2$ can always be formally mapped into a barcode. For instance, $C=256$ corresponds to an 8-bit grey scale and each bit can be 
represented as a binary variable with two possible configurations (B or W, by convention). As such, images with $C>2$ can be mapped into a ``barcode'' image with a higher number of pixels.


The general problem of barcode discrimination can be depicted as in Fig.~\ref{fig:barcode}c and \ref{fig:barcode}d. 
According to our notation, each pixel of a barcode has two possible grey-levels $i\in\{B,W\}$ and therefore corresponds to two possible quantum channels $\mathcal E_B$ and $\mathcal E_W$. For barcode decoding, we assume that the pixels are independently probed, so that the input state takes the tensor-product form $\tilde \rho =\rho_0^{\otimes n}$. Note that this assumption does not reduce the generality of our treatment. In fact, for the quantum source this leads to one of the best possible choices (tensor product of TMSV states). For the classical source, we know that independent and identical coherent states are able to saturate the lower bounds for general mixtures of multi-mode coherent states~\cite{zhuang2020entanglement}.
As for detection, the general scheme to correctly distinguish the various configurations consists in choosing a mixing
optical circuit, followed by measurements and classical post-processing algorithms 
as in Fig.~\ref{fig:barcode}d.
From an operational point of view, a suboptimal solution can be found for this problem by restricting to a cascade of beam splitters and phase shifters with tunable parameters, followed by independent measurements 
(e.g. homodyne or photodetection), similar to that of Ref.~\cite{zhuang2019physical}; while for the classical post-processing we may employ statistical classification 
algorithms commonly employed in machine learning applications, e.g. based on neural networks \cite{murphy2012machine}.
The suboptimal solution is then numerically investigated by minimizing the parameters of the optical and neural 
networks in order to minimize $p_{\rm err}$. 
When photodetection measurements are employed, analytic gradients can be computed following Ref.~\cite{banchiGBS} to 
speed-up the optimization algorithm. 
In this paper however we focus on the most general case and study the fundamental limits of barcode 
decoding and patter recognition, 
introducing different theoretical limits that any possible scheme must satisfy. 
Indeed, the physical optical circuit and measurements, and also  the classical post-processing algorithm 
in Fig.~\ref{fig:barcode}d, can all be reabsorbed into an abstract POVM that must be optimized.

We start by considering the case where a barcode with $n$ pixels is prepared in one of all possible $2^n$ patterns, each with equal prior. Then, we will consider the case where the patterns are restricted to specific configurations, where $k$ white pixels are randomly positioned within a grid of otherwise black pixels.





Starting from the input state
$\tilde \rho =\rho_0^{\otimes n}$, the possible states at the output of the barcode 
$\rho_{\bs i} = \mathcal I^n 
	\otimes \mathcal E_{i_1}\otimes\cdots\otimes\mathcal E_{i_n} (\tilde \rho)$,
take the product form 
\begin{equation}
	\rho_{\bs i} = \bigotimes_{k=1}^n \rho_{i_k}.
	\label{RhoIdlers}
\end{equation}
Correspondingly, the fidelities can be simplified as 
\begin{equation}
	F(\rho_{\bs i},\rho_{\bs j}) = F(\rho_W,\rho_B)^{{\rm hamming}(\bs{i},\bs{j})},
	\label{FidIdlers}
\end{equation}
where $\rho_i = \mathcal I\otimes\mathcal E_i(\rho_{0})$ for $i\in\{B,W\}$ and 
${\rm hamming}(\bs{i},\bs{j})$ is the Hamming distance between the two binary images 
${\bs i}$ and ${\bs j}$,
namely the number of pixels in which the two images differ. 
Using the properties of the Hamming distance, in Appendix~\ref{a:bound} we show that, for uniform 
{\it a priori} probabilities, the $M$-probing bounds~\eqref{BarnumMontanaro} become 
\begin{equation}
	\frac{(F^{2M}_{\rm max}+1)^n-1}{2^{n+1}} \leq 
	p_{\rm err} \leq 
	1-\left(1-\frac{F^M_{\rm max}}2\right)^n,
	\label{perrcomb}
\end{equation}
where $F_{\rm max}$ is the fidelity between any 
two (different) images with minimum Hamming distance [cf. Eq.~\eqref{FidIdlers}].

The minimum Hamming distance
is achieved when the two images differ by a single pixel. 
Thus, we get $F_{\rm max} = F(\rho_B,\rho_W)$, namely the maximum fidelity between any two images is given by the fidelity between the 
states describing the grey-levels of a single pixel. 
Using the Bernoulli's inequality, we then simplify Eq.~\eqref{perrcomb} as 
\begin{equation}
\frac{n}{2^{n+1}} F_{\rm max}^{2M} \leq	
p_{\rm err} \leq \frac n2 F_{\rm max}^M~.
\label{perrsimple}
\end{equation}
In Appendix~\ref{a:locmeas}, we show that the upper bound can be achieved using {\it local} measurements, 
namely where each pixel is measured independently from the others and 
$\Pi_{\bs i} = \bigotimes_{j=0}^{n-1} \Pi_{i_j}$ in Eq.~\eqref{measupattern}, though each 
pixel and its respective idler may be measured together (see Fig.~\ref{fig:barcode}e). 
Once we restrict to local operations, the optimum is achieved by independent
Helstrom measurements \cite{helstrom1969quantum} and the upper bound in Eq.~\eqref{perrsimple}
follows from Fuchs–van de Graaf inequalities \cite{fuchs1999cryptographic}.
A sub-optimal local measurement is obtained by combining the signal and idler via 
a beam splitter followed by independent measurements 
\cite{Qreading}.
Moreover, in the supplementary material \cite{suppmat} we also discuss different inequalities on
$p_{\rm err}$ based on the multiple quantum Chernoff bound~\cite{QSDReview,MultiChernoff,GaussianQCB2008}.

Two interesting observations can be made from the bounds~\eqref{perrsimple}.
First, the upper bound for the error probability becomes small whenever
$	n F^M(\rho_W,\rho_B) \ll 1$.
This implies that, although the set of images (namely barcode 
configurations) grows exponentially with the number of pixels as $2^n$, the required fidelities 
to accurately distinguish all configurations should decrease polynomially with $1/n$. 
In particular, $M=\mathcal O(\log n)$ copies are needed for correct discrimination. 
The second observation is that, due the factor $2^{-n}$, 
the lower bound in Eq.~\eqref{perrsimple} decreases exponentially with $n$. 
As we show in Appendix~\ref{a:locmeas}, this factor disappears from the lower bound 
when local measurements are employed.  
It is known that, in general,
optimum mixed state discrimination requires a joint
measurement~\cite{calsamiglia2010,bandyopadhyay2011}, yet
in our setting  optimal global measurements may in principle 
exponentially reduce the probability of error. 
Nonetheless, it is currently an open question to verify whether and exponentially decreasing 
error is achievable with optimal quantum measurements. In the next section we will claim 
quantum advantage whenever the upper bound on $p_{\rm err}$ obtained with entangled states and local 
measurements is smaller than the lower bound on $p_{\rm err}$ obtained with classical states and possibly global 
measurements, as schematically shown in Figs.~\ref{fig:barcode}d)~\ref{fig:barcode}e). 
Therefore, 
if the lower bound in \eqref{perrcomb} is loose, the regimes for quantum advantage are larger.

In the previous bounds we considered a uniform distribution of black an white pixels 
in the barcode. We may also consider a different encoding with a fixed 
number of white pixels, generalizing the results of Ref.~\cite{zhuang2020entanglement}. 
The task is then to find the position of $k$ white pixels in a barcode with $n$ bars. 
The number of possible configurations is $\binom nk\approx 2^{n H(k/n)}$ where 
$H$ is the binary entropy function and the approximation holds when both $n$ and $k$ are large. 
Therefore, in that regime, the configuration space grows exponentially with $n$, as in the 
uniform case discussed above. In the asymptotic regime we obtain the following bounds 
\begin{equation}
	\frac{k(n-k)}{2^{nH(k/n)+1}} F_{\rm max}^{4M} \lesssim
	p^{k-{\rm whites}}_{\rm err} \lesssim
	k(n-k) F_{\rm max}^{2M},
	\label{kCPF}
\end{equation}
while the exact expressions for finite $M$, $n$ and $k$ are discussed in Appendix~\ref{a:kcpf}.

\subsection{Quantum enhancement}\label{s:enhancement}
We now focus on photonic setups and model each pixel as a bosonic channel with 
transmissivity $\eta_i$, where $i\in\{B,W\}$ is the pixel color. 
We discuss the regime where we get an advantage from using entangled photons as input. 
We compare the case where each input $\rho_0$ is a TMSV state $\ket{\Phi_{N_S}}$ 
with $N_S$ average photons and the case where the input is a coherent state with the same 
number of signal photons $\ket{\sqrt{N_S}}\otimes \ket{0}$ (where the vacuum state means that no idler is used).
Note that one can replace the vacuum idler with an arbitrary state, such as a strong local oscillator in a coherent state, however that will not give a better performance when the optimum measurement is considered.
Assuming $M$ probings of the barcode, we have a total of $N_{\rm tot} = MN_S$ mean photons irradiated over each pixel. According to the analysis from the previous section, provable quantum advantage can be achieved
whenever the upper bound from Ineqs.~\eqref{perrsimple}, obtained with TMSV input states, is less than the lower bound obtained with coherent state inputs. Since the upper bound in~\eqref{perrsimple} is obtained with local measurements, what we call ``provable advantage'' 
means that possibly non-optimal local measurement strategies with entangled inputs 
beat any strategy with coherent states, even when the latter is enhanced by
complex global measurements. 
Provable quantum advantage may be more difficult for larger $n$, given 
the exponentially decreasing factor in the lower bound of Eq.~\eqref{perrsimple}, 
but here we show that it can be achieved for every number of pixels $n$ with suitably large number of probings $M$.

Using the formula for the 
fidelity between two generally-mixed Gaussian states \cite{Fidelity,marian2012uhlmann}, for TMSV states at the input, we compute (see Appendix~\ref{a:fid})
\begin{equation}
	F_q(\rho_W,\rho_B)^M = 
	\left(\frac{1}{1+N_S\Delta_q}\right)^M \geq e^{-M N_S \Delta_{\rm q}},
	\label{fidelityQ}
\end{equation}
where the index $q$ stands for {\it quantum} and 
\begin{equation}
	\Delta_{\rm q} = 	1- \sqrt{(1-\eta_W) (1-\eta_B)}-\sqrt{\eta_W \eta_B}.
\end{equation}
For a coherent-state input, we instead have
\begin{align}
	F_{\rm c}(\rho_W,\rho_B)^M &= 
	e^{-M N_S \Delta_{\rm c}},
&
\Delta_{\rm c} &= \frac{(\sqrt{\eta_B}-\sqrt{\eta_W})^2}2.
\label{fidelityC}
\end{align}
By comparing Eqs.~\eqref{fidelityQ} and \eqref{fidelityC} we see that, for fixed $M$, the fidelity between coherent states 
displays an exponential decay as a function of $N_S$, while for 
quantum states we see a polynomial decay in $N_S$. 
Nonetheless, 
for large $M$ and small $N_S$, the inequality in Eq.~\eqref{fidelityQ} becomes tight and, 
since $\Delta_{\rm q} \geq \Delta_{\rm c}$, in that limit
we find that quantum light always provides an advantage
for discrimination, irrespective of the values of $\eta_W$ and $\eta_B$. 
The limits of small $N_S$ and large $M$ are widely employed to show quantum advantage 
and can be realized experimentally with little imperfections \cite{zhang2013entanglement}.
Therefore, from now on  we will focus on such limits, $M\to\infty$ and $N_S\to0$, while keeping
fixed the total mean number of photons $M N_S$ irradiated over each pixel.

To properly demonstrate the advantage, we need to 
show that the upper bound on the probability of error using quantum light is smaller than the 
lower bound on the probability of error using coherent states. From Ineqs.~\eqref{perrsimple}, we see that this happens when 
$	F_{\rm c}^{2M} \geq 2^n F_{\rm q}^M$.
Setting $ n =\nu M N_S$, the previous inequality implies that quantum advantage is obtained for 
\begin{equation}
	\nu\leq  \nu_{\rm th} =\frac{\Delta_{\rm q}-2\Delta_{\rm c}}{\log 2},
	\label{nuthreshold}
\end{equation}
which is a barcode multi-pixel generalization of the ``threshold energy'' theorem proven in the context of single-cell quantum reading~\cite{Qreading}. 

\begin{figure}[t]
	\centering
	\includegraphics[width=0.99\linewidth]{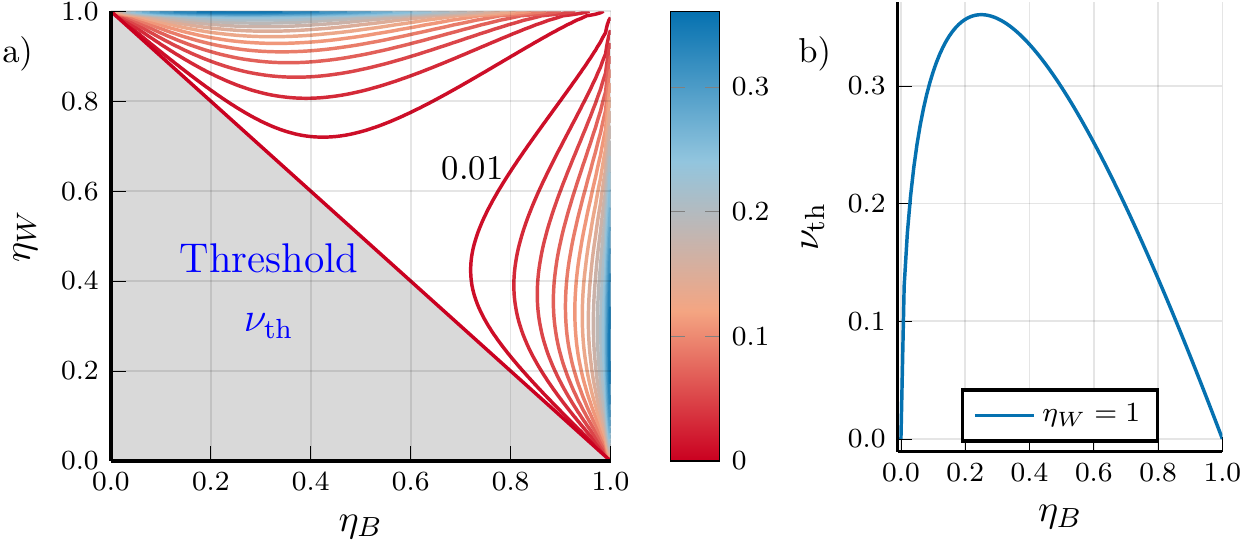}
	\caption{{ Regimes of provable quantum advantage}. a)
		Threshold value from Eq.~\eqref{nuthreshold} as a function 
		of $\eta_W$ and $\eta_B$. The threshold $\nu_{\rm th}$ is negative in 
		the filled gray area and positive for $\eta_W>1-\eta_B$. Contours 
		are from 0.01 in steps of 0.02. b) Threshold $\nu_{\rm th}$ for 
		$\eta_W=1$. 
		Whenever $n \leq M N_S \nu_{\rm th}$, or similarly $M\geq \frac{n}{N_s\nu_{\rm th}}$, 
		entangled light beats classical strategies based on coherent states. 
	}%
	\label{fig:nuth}
\end{figure}

According to Eq.~(\ref{nuthreshold}), whenever the number $n$ of pixels is smaller than a certain threshold, entangled light 
always provides an advantage in the discrimination of barcode configurations (barcode decoding) with respect to the best classical strategy with the same signal energy, even when 
the latter uses possibly complex global measurements. The behaviour of $\nu_{\rm th}$ as a function 
of $\eta_W$ and $\eta_B$ is numerically shown in Fig.~\ref{fig:nuth}. 

Quantum advantage can also be proven when we consider a prior distribution for the barcode configurations that is non-uniform, more precisely for the case where the number $k$ of white pixels is fixed. Using Ineqs.~\eqref{kCPF}, we find that there is a provable quantum advantage when 
$	F_{\rm c}^{4M} \geq 2^{n H(k/n)+1} F_{\rm q}^{2M}$, namely when 
$n H(k/n)+1 \leq 2\nu_{\rm th} M N_S$. Therefore, as in the previous case, quantum advantage 
may be observed when the number of pixels is sufficiently small or the number of probes $M$ 
is sufficiently large, as long as $\nu_{\rm th}\geq 0$.

It is currently an open question to prove whether or not the lower bound in~\eqref{perrsimple} can be achieved when classical light is employed. Nonetheless, our 
	analysis shows that even assuming that such bound can be achieved with classical inputs, 
	a strategy based on entangled light and the much simpler local measurements can beat 
	any approach based on coherent states. 
	On the other hand, if only local measurements can be performed, then the factor $2^{-n}$ in 
	the lower bound ~\eqref{perrsimple} disappears (see Appendix~\ref{a:locmeas}). This corresponds 
	to the case $\nu=0$. Therefore, in that case, whenever $\nu_{\rm th} > 0$,
	namely when $\eta_W>1-\eta_B$, 
quantum light provides an advantage for decoding uniformly-distributed barcodes, irrespective of $n$.



\section{Pattern recognition}\label{s:pattern}

\subsection{Statistical pattern classification}

\begin{figure}[ht!]
	\centering
	\includegraphics[width=0.9\linewidth]{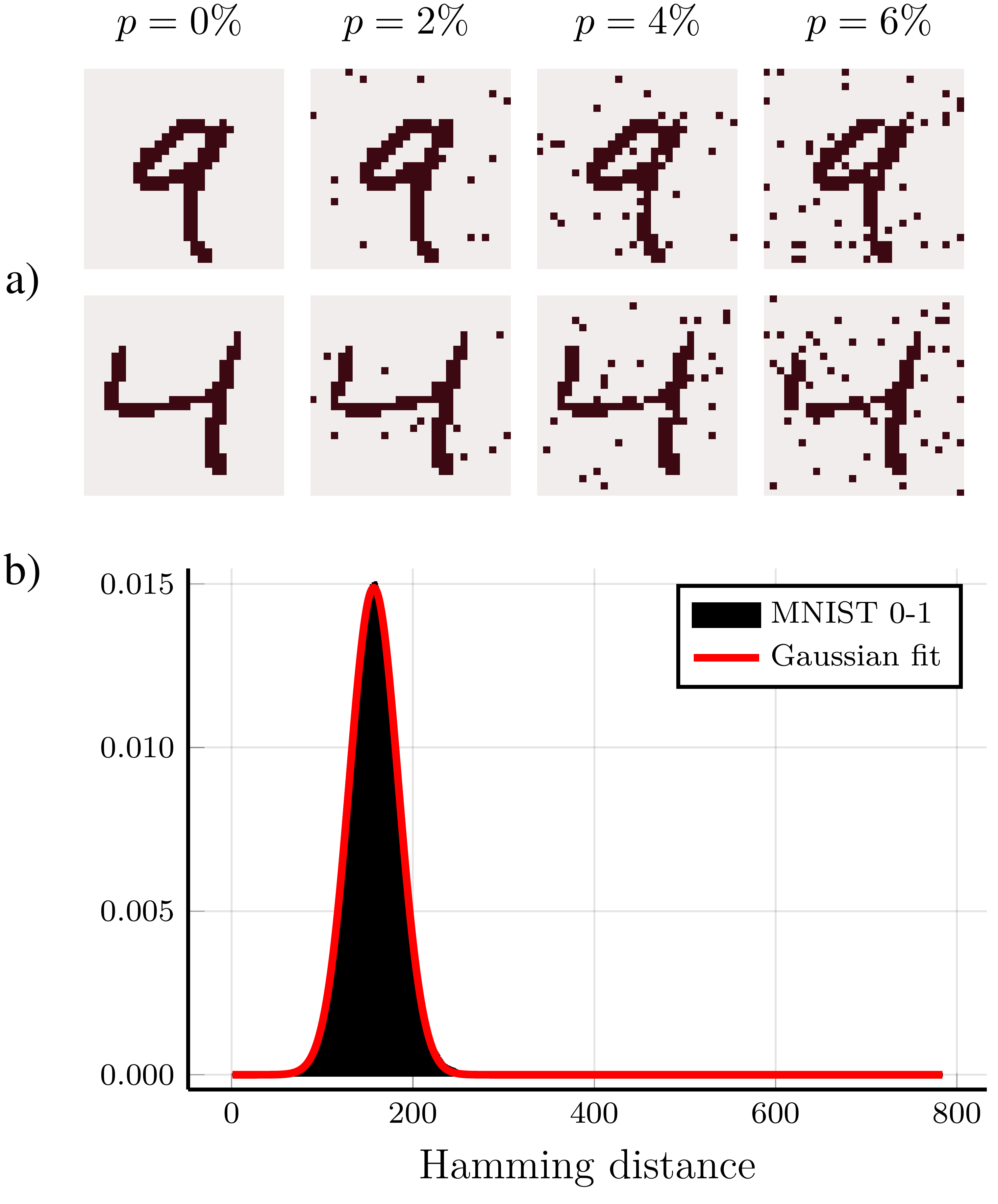}
	\caption{{ Pattern recognition.}
		(a) Images from the MNIST dataset without pixel recognition error ($p=0\%$) 
		and with pixel error probabilities $p=2\%,4\%,6\%$, where each pixel is randomly flipped with probability $p$. 
		(b) Probability $P_{cc'}(h)$ that one image from class $c$ has Hamming distance $h$, with  $0\leq h \leq 28{\times} 28$,
		from another image from class $c'$. The empirical histogram is evaluated for images from the MNIST dataset that 
		correspond to digits 0 and 1. The Gaussian fit has mean $\mu_{01} \simeq 157$ and standard deviation $\sigma\simeq 27$. 
		Different digits show a similar behaviour with $110\lesssim \mu_{cc'}\lesssim 167$, where the minimum is achieved 
		between 1 and 7.
	}%
	\label{fig:digits}
\end{figure}

We now focus on the problem of pattern recognition. 
Consider the problem of recognizing handwritten digits as shown in Fig.~\ref{fig:digits}a, whose images have been adapted from the MNIST dataset \cite{lecun1998gradient}. Each image depicts a single handwritten digit and the task is to 
extract from the image the corresponding number 0-9. From an algorithmic perspective, this task is more complex than the mere decision of whether 
a pixel is black or white but, from a physical point of view, this problem is actually simpler 
as errors are tolerated. Indeed, a human is able to instantly recognize all the numbers
in Fig.~\ref{fig:digits}a even when some of the pixels are randomly flipped. Therefore, 
for reliable pattern recognition, it is not necessary to perfectly reconstruct the entire image. Compared to the barcode configurations of Fig.~\ref{fig:barcode}, where each pixel provides  important information, here the goal is to recognize a global property that is robust against individual pixel errors, which means that entirely different strategies are possible.


In statistical learning theory \cite{hastie2009elements},
different learning tasks, such as image classification, can be modeled using probabilities. 
We consider the abstract space of all possible images and define the probability $\pi_{\bs i}$ 
of getting the image $\bs i$ -- this is unknown and generally not uniform. 
Image classification is a rule that attaches a certain label $c$, or class, to a given image 
$\bs i$. If this rule is deterministic, then it can be modeled via a function $c=f(\bs i)$ but, more generally, the strategy is stochastic: given a certain image, the rule
predicts different possible classes $c$ with a probability distribution $P(c|\bs i)$.
Let us consider a pair $(c,\bs i)$ and assume that, given our data, we have built 
a classifier $\tilde c(\bs i)$ 
that assigns a certain class $\tilde c(\bs i)$ to the image $\bs i$. The error in our classification 
can be described by a loss matrix with elements $L_{c\tilde c}$ that models the error 
of misclassification. The common choice is the 0-1 loss 
with $L_{cc}=0$ and $L_{c\tilde c}=1$ for $\tilde c\neq c$. 
By using the conditioning $P(c,\bs i)=P(c|\bs i)\pi_{\bs i} = P(\bs i|c) P(c)$, the expected classification 
error can be written as
\begin{equation}
	E = \ave_{(c,\bs i)\sim P(c,\bs i)}[L_{c,\tilde c(\bs i)}] = 
	1-\sum_{\bs i}  P(\tilde c(\bs i)|\bs i)\,\pi_{\bs i}~.
	\label{experr}
\end{equation}
For known $P(c,\bs i)$ the optimal classifier is then the one minimizing the
expected classification error, 
$	\tilde c_{\rm B}(\bs i) = \argmax_c P(c|\bs i)$,
which is called {\it Bayes classifier}, while 
the resulting error from \eqref{experr},
is called {\it Bayes rate}. 
The Bayes rate represents the theoretical minimum error that can be expected with 
the optimal classifier.

We now study the error of pattern classification when images are noisy, for instance 
due to an imperfect detection. The setup is the same of Fig.~\ref{fig:barcode},
where light, either quantum or classical, is used to illuminate the pattern (e.g.~a 
handwritten digit as in Fig.~\ref{fig:digits}a) and, from the detected output, the task 
is to find the correct class (e.g.~a number between 0-9). 
For this purpose, we introduce the minimum error as a generalization of 
Eqs.~\eqref{experr} and \eqref{measupattern}
\begin{equation}
	E^{\rm Q} := 
	\min_{\{\Pi_c\}}  \sum_{c\neq \tilde c}\sum_{\bs i}   \Tr[	\Pi_{\tilde c}\rho_{\bs i}^{\otimes M}] P(c,\bs i)~,
	\label{Equantum}
\end{equation}
where the operators $\{\Pi_c\}$ define a POVM whose measurement outcome $c$ 
predicts the  class of the image $\bs i$ encoded into the quantum state $\rho_{\bs i}$. 
For a two-class decision problem, the optimal POVM can be explicitly found by  Helstrom theorem 
\cite{helstrom1969quantum}. When the number of classes is larger, a 
``pretty good'' approximation to the optimal POVM can be obtained with pretty good measurements. 
For general measurements, 
we may derive bounds similar to \eqref{BarnumMontanaro}, generalizing 
\cite{Barnum,Montanaro,montanaro2019pretty} (see Appendix~\ref{s:pgm} for details). 
\begin{align}
	B[F(\rho_B,\rho_W)^{2M}] \leq	E^{\rm Q} \leq 2K B[F(\rho_B,\rho_W)^M],
	\label{PatternBounds}
\end{align}
where $K$ 
is such that $K^{-1}$ is the minimum non-zero value of $\sqrt{P(c,\bs i)P(c',\bs i')}$,
which is independent of $M$, $\rho_W$ and $\rho_B$, and we have defined 
\begin{equation}
	B[F] := \frac12\sum_{c\neq c'} P(c)P(c') \sum_{h=1}^n P_{cc'}(h) F^h, 
	\label{BF}
\end{equation}
where $P_{cc'}(h)$ is the probability that two images from different classes $c$ and $c'$ have 
Hamming distance $h$. 
For large $M$, the term with minimum Hamming distance dominates and we may write
\begin{equation}
	B[F^M] \propto F^{M h_{\rm min}}~,
	\label{Bscaling}
\end{equation}
where $h_{\rm min}$ is the minimum Hamming distance between two images from different classes.
The Ineqs.~\eqref{PatternBounds} and the expansion \eqref{Bscaling} represent the most 
important results of this section, generalizing Ineqs.~\eqref{perrsimple} and \eqref{kCPF} 
to the problem of pattern recognition. By comparing those bounds, we find that 
quantum-enhanced pattern recognition is significantly simpler than barcode discrimination 
when $h_{\rm min}>1$, as the error decreases with the faster rate \eqref{Bscaling}.

The error $E^{\rm Q}$ is a quantum generalization of Bayes rate,
and quantifies the theoretical optimal performance of the classification rule.
However, alike the Bayes rate, it is difficult to compute since the distribution 
$P(c,\bs i)$ is typically unknown, and no closed-form solutions to \eqref{Equantum} 
exist beyond the two-class case. To solve these issues, 
in the next section we propose a supervised learning approach where 
an optimal classification measurement is estimated
from a collection of correctly classified data.

\subsection{Supervised quantum pattern recognition}\label{s:supervi}

In data driven approaches the task is to approximate the optimal classifier 
via a collection of already classified examples $(c^{\mathcal T}_k,\bs i^{\mathcal T}_k)$. 
The set $\mathcal T= \{(c^{\mathcal T}_k,\bs i^{\mathcal T}_k)~{\rm for}~k=1,\dots,T\}$ is
called  {\it training set} and $T$ is its cardinality. 
In the framework of statistical learning theory, we can treat 
the elements of this set as 
{\it samples} from the abstract and unknown joint probability distribution $P(c,\bs i)$ introduced above.
Then, in the limit of large $T$ we may approximate the averages with respect to $P(c,\bs i)$ with 
{\it empirical} averages over the training set. This allows us to explicitly compute the classification 
error \eqref{Equantum} and the theoretical bounds \eqref{PatternBounds}.
Therefore we define an empirical learning method, also called ``training'', as an 
optimization of the POVM $\{\Pi_c\}$ to correctly classify, as much as possible, the 
known samples from the training set $\mathcal T$ 
\begin{equation}
	{\rm training:~} 
	\min_{\{\Pi_c\}}  \frac1T \sum_{k=1}^T \sum_{c\neq c_k^\mathcal T }
	\Tr[	\Pi_{c}\rho_{\bs i_k^\mathcal T}^{\otimes M}] =: E^{\rm Q}_{\mathcal T}.
	\label{training}
\end{equation}
From an operational point of view, a suboptimal solution to optimal detection $\{\Pi_c\}$ 
can be found for instance 
as shown Fig.~\ref{fig:barcode}d and discussed in section \ref{s:barcode}, by optimizing over the 
available optical circuit, measurement schemes and classical post-processing. Here on the 
other hand we study the ultimate theoretical limits that any classification task must satisfy, 
studying the minimum training error $E^{\rm Q}_{\mathcal T}$ via bounds like \eqref{PatternBounds},
while the ability to classify unseen data will be discussed in the next section. 
Indeed, 
upper and lower bounds 
on $E^{\rm Q}_{\mathcal T}$ can be obtained (see Appendix~\ref{s:pgm})
as an average fidelity between states $\rho_{\bs i_k^\mathcal T}$ and $ \rho_{\bs i_{k'}^{\mathcal T}}$ 
whose images from the training set have different classes,  $c_k^{\mathcal T}\neq c_{k'}^{\mathcal T}$.
Thanks to Eq.~\eqref{FidIdlers} we finally get
\begin{align}
	B_{\mathcal T}[F(\rho_B,\rho_W)^{2M}] \leq	E^{\rm Q}_{\mathcal T} \leq 2T B_{\mathcal T}[F(\rho_B,\rho_W)^M],
	\label{PatternBounds1}
\end{align}
where we have defined 
\begin{equation}
	B_{\mathcal T}[F] = 	\sum_{k, k' : c_k^{\mathcal T}\neq c_{k'}^{\mathcal T} } 
	\frac{F^{{\rm hamming}({\bs i_k^\mathcal T},{\bs i_{k'}^{\mathcal T}})}}{2T^2}.
	\label{Bfunc}
\end{equation}
It is simple to show that $B_{\mathcal T}[F]$ is a particular case of $B[F]$
from Eq.~\eqref{BF} in which averages 
over the abstract distribution are substituted with averages over the empirical distribution. 
As such, we may rewrite $B_{\mathcal T}$ as in Eq.~\eqref{BF} and obtain the large-$M$ scaling
\eqref{Bscaling}. 


As a relevant example, we consider the problem of handwritten digit classification with the 
MNIST dataset \cite{lecun1998gradient}. 
The MNIST dataset is composed of a training set of $60000$ images and 
corresponding classes, and a testing set of $10000$ images and 
corresponding classes. Each original image is in grey scale and has $n=28{\times}28$ pixels. For simplicity we first map each pixel to either black or white, depending on the closest grey-level. 
In this way, every image can be seen as a 2D barcode. For the MNIST dataset 
we see from Fig.~\ref{fig:digits}b) that the probability $P_{cc'}(h)$ that two images from different classes have 
Hamming distance $h$ resembles a Gaussian distribution with mean $\mu_{cc'}$ 
and standard deviation $\sigma_{cc'}$, and minimum non-zero value $h^{\rm min}_{cc'}$. 
Using this approximation, we find in Appendix~\ref{s:pgm} analytical approximations for $B_{\mathcal T}[F]$, 
recovering the scaling \eqref{Bscaling}, where $h_{\rm min} = \min_{c\neq c'} h^{\rm min}_{cc'}$. 
For the MNIST dataset, we find $h_{\rm min}=25$. 
Therefore, from \eqref{PatternBounds1} we may 
get an error that decays as   $E^{\rm Q}_{\mathcal T} \approx F(\rho_B,\rho_W)^{\alpha M h_{\rm min}}$, 
independently on the number of pixels $n$ and with 
$1\leq \alpha\leq 2$. Moreover, thanks to Ineqs.~\eqref{PatternBounds1} we may define a guaranteed quantum 
advantage when the upper bound obtained with entangled states is smaller than the lower bound obtained with 
classical data, namely when $2T F_q^{M h_{\rm min}}\leq F_c^{2 M h_{\rm min}}$. Since the training set 
is normally very large, we may set $2T=2^{\nu M h_{\rm min} N_S}$ for some $\nu$ and the above inequality 
becomes equivalent to \eqref{nuthreshold}, in the limit $M\to \infty$ and $N_S\to0$. Therefore, 
we may repeat the same analysis of Sec.~\ref{s:enhancement}: 
 whenever $\nu_{\rm th}>0$ (see Fig.~\ref{fig:nuth}), quantum advantage can be proven 
for training sets whose dimension is bounded as $2T\leq 2^{\nu_{\rm th} M N_S h_{\rm min} }$.
In other terms, setting $N_{\rm tot}=MN_S$ we find a simple relation between the number of photons to show 
quantum advantage and the dimension of the training set as 
\begin{equation}
	N_{\rm tot}  \geq \frac{\log_2(2T)}{\nu_{\rm th}h_{\rm min}} \simeq 0.65\, \nu_{\rm th}^{-1} ~.
\end{equation}
In the above expression the first inequality holds in general, while the approximated numerical 
value is for the MNIST dataset, where $h_{\rm min}=25$ and $T=6{\times} 10^4$.

To conclude this section we note that 
unlike \eqref{perrcomb}, the upper bound in \eqref{PatternBounds1} is achieved with 
global measurements, so a strategy like the one in Fig.~\ref{fig:barcode}f may be needed to achieve 
such classification accuracy.
Bounds with local measurement errors  are discussed in the 
next section,  where each pixel is detected independently.

\subsection{Independent on-pixel measurements}
In the previous section we have studied the ultimate physical limits for pattern recognition 
by optimizing over all the elements of the optical apparatus, namely the optical circuit, the 
measurements and the classical post-processing routines (Fig.~\ref{fig:barcode}c). Together these can all be 
described as an abstract global POVM, as in Eq.~\eqref{Equantum}. 
Here we consider a simplified setup, similar to that of Fig.~\ref{fig:barcode}c but without the optical 
circuit and with local measurements $\Pi_{\bs i}=\prod_{j=1}^N \Pi_{i_j}$. 
Here a noisy image is reconstructed first, and then a classical algorithm is used to classify it. 
As before, we call $\bs i$ the real physical configuration of the $n$ pixels, 
each either black or white $i_j=\{B,W\}$, and $\tilde{\bs i}$ the binary variables corresponding 
to the reconstructed image, read by the sensors. 
Using $M$ copies to 
perform the detection, all possible reconstructed images can appear with
probability $	p_{\rm read}(\tilde {\bs i}|\bs i) $ as in Eq.~\eqref{measupattern}. 
Considering also the classical classification routine, the local setup 
consists in choosing a non-optimal POVM in Eqs.~\eqref{Equantum} or \eqref{training} as 
\begin{equation}
	\Pi_c = \sum_{\bs i} A(c|\tilde{\bs i}) \prod_{j=1}^n \Pi_{\tilde{i}_j}~,
	\label{localalgo}
\end{equation}
where $A(c|\bs i)$ is any reliable (possibly non-linear) 
machine learning algorithm that can classify the reconstructed 
images. The above equation defines a POVM as long as $\sum_c A(c|\bs i)=1$ for all $\bs i$, which 
is an obvious requirement since every image must be in at least one class. 

The classical algorithm must be noise resilient, because some pixels 
might not be properly reconstructed, see e.g. Fig.~\ref{fig:digits}a.
Noise naturally occurs in
readouts that are made in reflection where the light is diffused back to the receiver.
Classification in the presence of different forms of noise has a large literature 
in machine learning \cite{angluin1988learning}. Here, we
assume that our training set is composed of noiseless 
images that are correctly classified,
namely that 
$c_k^{\mathcal T}$ is 
the true class of $i_k^{\mathcal T}$. 
Although not explicitly discussed here, 
it is possible to extend our analysis to noisy training sets via the method of importance reweighting
\cite{liu2015classification,aslam1996sample,manwani2013noise}.

As for the classical algorithm in Eq.~\eqref{localalgo}, 
there are different strategies to define a classifier given the training set,
all with different performances and ranges of applicability \cite{hastie2009elements}. 
Here we focus on the nearest neighbor classifier \cite{cover1967nearest}, defined as 
\begin{align}
	\tilde c^{\mathcal T}_{\rm NN}(\bs i)&= c^{\mathcal T}_{k_{\rm min}}~,
									&
	k_{\rm min} &= \argmin_k D(\bs i,\bs i_k^{\mathcal T})~,
	\label{nnrule}
\end{align}
where $D(\bs i,\bs i')$ is a suitable distance between two images. In other terms, 
classification of an unknown image $\bs i$ is done by selecting the class 
$c_{k_{\rm min}}^{\mathcal T}$ of the image from the training set that 
is closest to $\bs i$, according to distance $D$. 
The corresponding algorithm in \eqref{localalgo} is 
$A(c|\bs i) = \delta_{c,\tilde c^{\mathcal T}_{\rm NN}(\bs i)}$.
More advanced neural-network based algorithms will be considered 
in another paper \cite{cillian}.

In spite of being very simple, the nearest neighbor classifier 
has many desirable features. Indeed, under mild conditions, it has 
been proven \cite{cover1967nearest} that, for $T\to\infty$, 
the classification error using the nearest neighbor classifier 
is at most twice the Bayes rate, irrespective of the number 
of classes. More details are shown in the Supplemementary material~\cite{suppmat}, where we study the
performance of this classifier for finite $T$, i.e., for finite training sets.
Another feature is the ability to choose the most appropriate distance $D$. 
Here we choose the Hamming distance, which allows us to exploit many results from
previous sections.

In this section we consider quantum sources and sensors, but classical algorithms for nearest neighbor 
classification. Quantum computers can perform
nearest neighbor classification quicker than any classical counterpart \cite{wiebe2015quantum}, but 
how to mix those quantum algorithms with optical detection schemes is still an open problem.

Inserting Eq.~\eqref{localalgo} into \eqref{Equantum} 
and employing the nearest neighbor classifier 
we get 
	\begin{equation}
E^{\rm NN}_{\mathcal T} := 
\sum_{\bs i,\tilde{\bs i}} \sum_{c\neq \tilde{c}_{\rm NN}^{\mathcal T}(\tilde{\bs i})} P(c,\bs i)
\prod_{j=1}^n \Tr[	\Pi_{\tilde{i}_j}\rho_{i_j}^{\otimes M} ],
\label{Ennt}
\end{equation}
To understand this error, suppose that the pixel error probability $p$ is independent on 
whether the pixel is black or white. In this case, the probability that the reconstructed 
image $\tilde{\bs i}$ differs from the true image $\bs i$ in 
$k={\rm hamming}(\bs i,\tilde{\bs i})$ pixels is a binomial distribution $\propto p^k(1-p)^{n-k}$,
with mean $np$. Thanks to the analysis shown in Fig.~\ref{fig:digits}b we know that,
on average, as long as the number of wrongly detected pixels is smaller than the typical 
separation in Hamming distance between different classes, the nearest neighbour classifier 
should provide the correct result. For the transformed MNIST dataset, 
$n=28{\times}28$ and the typical 
number of flips between different classes is $\approx 160$ (see Fig.~\ref{fig:digits}b), so 
a pixel error probability up to $p\simeq 160/784\simeq 20\%$ should be tolerated by the algorithm.

\begin{figure}[t!]
	\centering
	\includegraphics[width=0.9\linewidth]{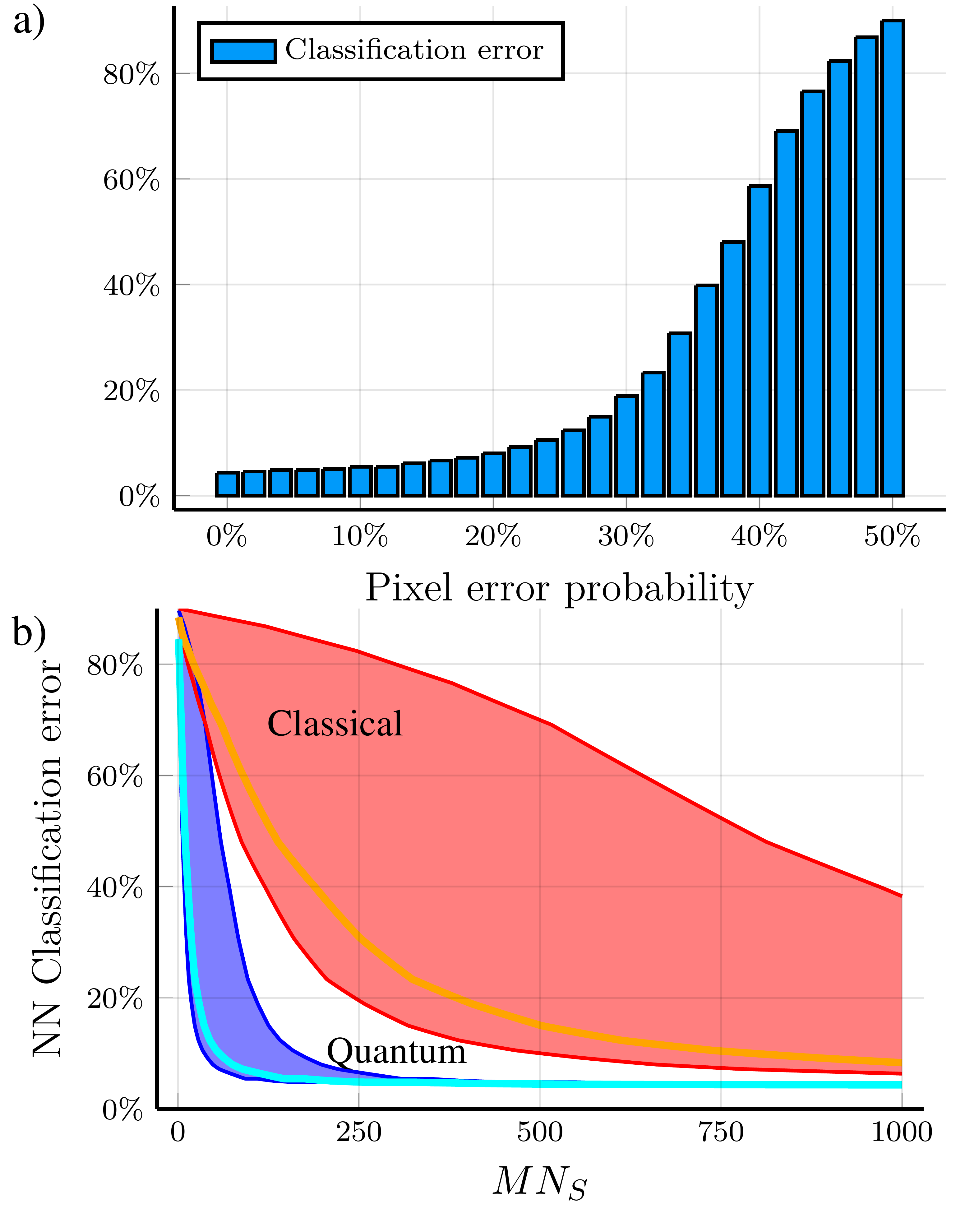}
	\caption{{ Pattern recognition with independent on-pixel measurements.}
		(a) Classification error, namely empirical probability of recognizing the wrong digit 
		using the nearest neighbor classifier with Hamming distance, as a function of the pixel error probability. 
		(b) Classification error when each pixel is probed using either coherent inputs or 
		entangled TMSV states. 
		The plot is generated 
		by combining the error coming from the single pixel error probability
		(see Appendix~\ref{a:locmeas})  with
		the classification error from (a). 
		We focus on the limit $M\to\infty$, $N_S\to0$ while keeping fixed the total mean number of photons $M N_S$ irradiated over each pixel. 
		The colored areas represent the region between the upper and lower bounds assuming quantum (blue) or classical (red) sources combined with optimal local measurements. These bounds depend on the quantum and classical fidelities from Eqs.~\eqref{fidelityQ} and~\eqref{fidelityC}. 
		The cyan and orange lines represents the performance with 
		quantum light (cyan) or classical coherent states (orange) using (non-optimal) photodetection measurements. 
		We set $\eta_B=0.9$ and $\eta_W=0.95$. 
	}%
	\label{fig:digits2}
\end{figure}

In Fig.~\ref{fig:digits2} we study the robustness of the nearest neighbor classifier via a numerical 
analysis with the transformed MNIST dataset, where each image 
is transformed into a 2D barcode as described in Sec.~\ref{s:supervi}.
We use such transformed training set to build a nearest neighbor classifier, 
and then estimate the error \eqref{Ennt} as an average over the testing set, namely 
as $N_{\rm wrong}/10000$ where 
$N_{\rm wrong}$ is the number of times that 
in the 10000 entries of the testing set, the predicted digit is different from the true one. 
Since the images from the testing set are samples from the 
abstract and unknown probability $P(c,\bs i)$ 
in the limit of infinitely large testing sets such estimate
converges to $E^{\rm NN}_{\mathcal T}$ from Eq.~\eqref{Ennt}. 
Moreover, since the images 
from the testing set are different from the ones in the training set, this error contains 
two terms: an error due to imperfect detection and an generalization error, since we 
are classifying previously unseen data. 

In Fig.~\ref{fig:digits2}a we study the classification error as a function of the 
probability $p$ of wrong pixel detection. As we see,
even for noiseless images, namely when $p=0$,
the classification error is still non-zero, as the nearest neighbor classifier may 
provide wrong outcomes. Nonetheless, as predicted, 
Fig.~\ref{fig:digits2}a shows that the nearest neighbor classifier is remarkably robust against 
relatively high pixel error probabilities $p$.

In Fig.~\ref{fig:digits2}b we combine the bound on the pixel error probability 
(see Eq.~\eqref{perrlocal}, for a single pixel $n=1$)  with the theoretical curve that 
predicts the classification error from the pixel error probability in Fig.~\ref{fig:digits2}a.
The bounds on the pixel error probabilities are obtained from the fidelities, 
Eqs.~\eqref{fidelityQ} and~\eqref{fidelityC}, which consider either coherent 
states or entangled TMSV states with the same average number of photons $MN_S$. 
The results from Fig.~\ref{fig:digits2}b show that the classification
error when we use quantum light is lower than the corresponding classical value. 
These results are based on the assumption 
that the detector performs the optimal Helstrom measurement, which 
may be complex to implement experimentally. Therefore, 
in Fig.~\ref{fig:digits2}b, we also 
consider the simpler photodetection measurement, where the POVM in Eq.~\eqref{measupattern} 
is a projection onto the Fock basis. The resulting pixel error probabilities 
with both coherent states and TMSV inputs are studied in the Supplementary Material~\cite{suppmat},
adapting the analysis from Ref.~\cite{ortolano2020experimental}. 
We see that even for this non-optimal measurement, entangled inputs always provide an 
advantage against purely classical coherent states for all possible values of $MN_S$. 
This advantage can be experimentally observed via a setup like that of Ref.~\cite{ortolano2020experimental}.

\section{Discussion}\label{s:conclusions}
In this work, we have investigated multi-pixel problems of quantum channel
discrimination, namely the identification of barcode configurations (equivalent
to readout of the stored data) and the classification of black and white
patterns, e.g.~given by noisy digital images of handwritten digits. In both
cases, we have shown that the use of quantum light based on entangled states
clearly outperforms classical strategies based on coherent states. 

For both quantum-enhanced 
barcode decoding and pattern recognition, we have
analytically studied, via bounds, the physical limits to the classification error that we may get by optimizing 
over all optical elements, measurements and classical post-processing. 
This allows us to to derive  explicit analytical conditions for the quantum advantage to hold. 
Moreover, 
the analysis of our bounds allows us to rigorously prove that quantum-enhanced 
pattern recognition can vastly reduce the classification error with respect to the mere 
independent measurement of each pixel. 

Nonetheless, being easier from the experimental point of view, 
we have also 
considered a simplified setup where all pixels are probed independently and,
for the problem of pattern recognition, we found that 
photon counting measurements are sufficient to show quantum advantage,
paving the way for an experimental demonstration with state of the art quantum
technology.

\acknowledgements

L.B.  acknowledges  support  by  the  program ``Rita  Levi  Montalcini''  for  young  researchers. 
Q.Z. acknowledges support from the Defense Advanced Research Projects Agency (DARPA) under Young Faculty Award (YFA) Grant No. N660012014029 and Craig M. Berge Dean's Faculty Fellowship of University
of Arizona.
S.P. acknowledges funding from EU Horizon 2020 Research and Innovation Action under grant agreement No. 862644 (Quantum Readout Techniques and Technologies, QUARTET).

\appendix

\section{Bounds on barcode discrimination error}\label{a:bound}
Consider a barcode with $n$ pixels or bars, where 
each bar can assume two possible values (black or white) as in Fig.~\ref{fig:barcode}. 
There are in total $2^n$ possible configurations and the discriminator 
must be able to  correctly identify each code. 
Using uniform a priori probabilities $\pi_{\bs i}=2^{-n}$ 
in \eqref{BarnumMontanaro} we get 
\begin{align}
	p_{\rm err} &\geq 2^{-(2n+1)} \sum_{\bs{i}\neq \bs{j}} F^{2M}(\rho_{\bs i},\rho_{\bs j}),
\label{Fidelity_LB}
\\
	p_{\rm err} &\leq 2^{-n}
	\sum_{{\bs i}\neq {\bs j}}F^{M}(\rho_{\bs i},\rho_{\bs j}).
\label{Fidelity_UB}
\end{align}
For some specific choices of the input, e.g. for coherent states or TMSV states, the states take the 
product form \eqref{RhoIdlers} and the fidelities can be written as \eqref{FidIdlers},
where ${\rm hamming}(\bs{i},\bs{j})$ is the Hamming distance between the two images. 
In order to find computable bounds for $p_{\rm err}$ we study then the following quantity
\begin{equation}
	D_n(f) = \frac1{2^n}\sum_{{\bs i}\neq {\bs j}} f^{{\rm hamming}(\bs{i},\bs{j})}.
	\label{Dnf}
\end{equation}
We find a closed form for $D_n(f)$ by recursion. We may write $\bs{i}=\{i_1,\bs{i'}\}$ 
where $\bs{i'}$ has $n-1$ components and similarly for $\bs j$. Then 
in \eqref{Dnf} we may separately consider all four possible values of $i_1$ and $j_1$,
noting that when $i_1\neq j_1$ we may also have $\bs{i'}=\bs{j'}$. 
Therefore we get 
\begin{align}
	D_n &= \frac1{2^n}\sum_{{\bs i}\neq {\bs j}}  
	f^{{\rm hamming}(i_1,j_1)}
	f^{{\rm hamming}(\bs{i'},\bs{j'})} \\&= 
	\frac{D_{n-1}+D_{n-1}+ 2f(D_{n-1}+2^{n-1}/2^{n-1}) }2
																						 \\&=
	(1+f)D_{n-1}+ f~,
\end{align}
where each term in the sum corresponds to $(i_1,j_1) = \{(0,0),(1,1),(1,0),(0,1)\}.$
The solution to the above recursion is 
\begin{equation}
	D_n(f) = (f+1)^n-1~,
\end{equation}
from which we get the following inequalities 
\begin{align}
	p_{\rm err} &\geq \frac{[F^{2M}(\rho_W,\rho_B)+1]^n-1}{2^{n+1}},
	\label{barcodeLB}
\\
	p_{\rm err} &\leq 
[F^{M}(\rho_W,\rho_B)+1]^n-1
	\nonumber \leq \\& \leq 
	e^{n F^{M}(\rho_W,\rho_B)}-1,
	\label{barcodeUB}
\end{align}
where in the last line we use the inequality $(1+x)^r\leq e^{rx}$.
A small error $p_{\rm err}\leq \epsilon$ is then obtained 
if we use $M$ copies with
\begin{equation}
	M \geq \frac{-\log{\frac{\log(1+\epsilon)}n}}{-\log F}.
\end{equation}

Finally we study the asymptotic performance for $M\to\infty$. 
From the definition of the state \eqref{RhoIdlers} 
we may write 
\begin{equation}
	F_{\rm max} = F(\rho_W,\rho_B)^{ \min_{i\neq j}{\rm hamming}(\bs{i},\bs{j})}
	= F(\rho_W,\rho_B),
\end{equation}
and taking the limit in 
Eqs.\eqref{barcodeLB}-\eqref{barcodeUB} we get 
\begin{equation}
	\frac{n}{2^{n+1}}F^{2M} \lesssim p_{\rm err} \lesssim n F^M,
	\label{asympbound}
\end{equation}
which is valid for large $M$.

\section{Local measurements}\label{a:locmeas}
When local measurements are employed together with product states as in \eqref{RhoIdlers}, then the success 
probability can be simplified as the probability of independently discriminating each pixel, namely 
\begin{equation}
p^{\rm local}_{\rm succ} = (p^{\rm pixel}_{\rm succ})^n,
\end{equation}
where $p^{\rm local}_{\rm succ}$ is the success probability of discriminating each image \eqref{success} 
using local measurements only and $p_{\rm succ}^{\rm pixel}$ is the 
success probability of detecting the grey-level of a single pixel. 
For equal {\it a priori} probabilities, the success probability can be written as 
$	p_{\rm succ}^{\rm pixel} = (\Tr [ \Pi_B \rho_B] + \Tr[\Pi_W \rho_W])/2 = 
\frac12 + \frac14 \Tr[(\Pi_B-\Pi_W)(\rho_B-\rho_W)]$ and, 
thanks to the Helstrom theorem 
\cite{helstrom1969quantum,bengtsson2017geometry},
minimum error is obtained 
when $\Pi_B$ and $\Pi_W$ are, respectively, the projectors onto the positive and 
negative subspace of $\rho_B-\rho_W$. Therefore, with optimal Helstrom measurements 
the success probability  is given by 
$p_{\rm succ}^{\rm pixel} = 1/2+1/4 \| \rho_W-\rho_B\|_1$.
Using the Fuchs–van de Graaf inequalities \cite{fuchs1999cryptographic},
\begin{equation}
1-{F(\rho,\sigma)} \le \frac12 \|\rho-\sigma\|_1 \le\sqrt{1-F(\rho,\sigma)^2},
\label{FvdG}
\end{equation}
and the fact that 
$F_{\rm max} = F(\rho_B,\rho_W)$  we find 
\begin{align}
	1-\frac{F_{\rm max}}2 &\leq  p_{\rm succ}^{\rm pixel} \leq \frac{1+\sqrt{1-F_{\rm max}^2}}2,
	\label{psuccpixel}
\\
	\frac{1-\sqrt{1-F_{\rm max}^2}}2 
												&\leq p_{\rm err}^{\rm pixel} \leq 
	\frac{F_{\rm max}}2 ~.
	\label{perrpixel}
\end{align}
From the above inequalities, we may get a bound on the error probability
$p_{\rm err}^{\rm local}=1-p_{\rm succ}^{\rm local}$
using local measurements and $M$ copies. The result is 
\begin{equation}
	1-\frac{(1+\sqrt{1-F_{\rm max}^{2M}})^n}{2^n}
	\leq p_{\rm err}^{\rm local} \leq
	1-\left(1-\frac{F^M_{\rm max}}2\right)^n 
	\label{perrlocal}
 .
\end{equation}
Performing the asymptotic analysis, for large $M$ we find 
\begin{equation}
	\frac{n}{4}F^{2M} \lesssim p_{\rm err} \lesssim \frac n2 F^M,
	\label{asympboundloc}
\end{equation}
which results in the same decaying rate of Eq.~\eqref{asympbound}. Therefore, for small $n$ (and large $M$), the use of global measurements 
does not increase our ability to distinguish the product states \eqref{RhoIdlers}. 
However, for large $n$ the factor $2^{-n-1}$ in \eqref{asympbound} shows 
that non-local quantum measurements might provide an important advantage.

We now show that the upper bound from Eq.~\eqref{perrlocal} is tighter than 
that of Eq.~\eqref{barcodeUB}, namely that
\begin{equation}
(F_{\rm max}^M+1)^n-1 \geq
	1-\left(1-\frac{F_{\rm max}^{M}}2\right)^n 
	~.
		\label{locineq}
\end{equation}
Indeed, for any $0\leq f \leq 1$ we find 
\begin{align}
	(f+1)^n& + (1-f/2)^n = \\&= 2 + \sum_{k=1}^n 
	\binom{n}{k} f^k[1 + (-1/2)^k] \geq 2~,
	\nonumber
\end{align}
given that all terms in the sum are positive. Substituting $f=F_{\rm max}^M$ in 
the above equation we get \eqref{locineq}. 
The upper bound from \eqref{perrlocal} is always tighter than \eqref{barcodeUB} and we may combine the 
two bounds \eqref{barcodeLB} and \eqref{perrlocal} as in Eq.~\eqref{perrcomb}. 
Finally, using the Bernoulli's inequality, 
$(1+x)^n \geq 1+nx$ when $x\geq -1$, we find Eq.~\eqref{perrsimple},
which is valid for any $n$ and $M$. The above bounds coincide with 
what we get from the asymptotic analysis,
Eqs.~\eqref{asympbound} and \eqref{asympboundloc}, showing that 
the Bernoulli's inequality is tight in that regime. 

\section{Non-uniform a priori probabilities}\label{a:kcpf}

Here we consider a different problem where the set of images or configurations is restricted to have 
only $k$ white pixels. We may also call this problem $k$ channel position finding, $k$-CPF, being a generalization of the idea of CPF introduced in Ref.~\cite{zhuang2020entanglement}. Namely, we know that there will be exactly $k$ target channels (white pixels), but we do not know their positions. There are $n$ choose $k$ configurations in this set. The upper and lower bounds in Eqs.~\eqref{Fidelity_LB} and~\eqref{Fidelity_UB} can be solved as follows. 

Similar to Eq.~\eqref{Dnf}, we can define 
\begin{align}
	D_n^k(f) &= \frac1{\binom nk} 
	\sum_{{\bs i}\neq {\bs j}, k-CPF} f^{{\rm hamming}(\bs{i},\bs{j})}, 
	\label{Dnf_k}
\end{align}
where the summation is over configurations $\bm i \neq \bm j$ with $k$ target channels, and we can express Eqs.~\eqref{Fidelity_LB} and~\eqref{Fidelity_UB} as
\begin{align}
	p^{k-{\rm CPF}}_{\rm err} &\geq 
	\frac{D_n^k[F(\rho_B,\rho_W)^{2M}]}{2 \binom nk},
	\label{PELBF}
\\
	p^{k-{\rm CPF}}_{\rm err} &\leq 
	D_n^k[F(\rho_B,\rho_W)^M].
	\label{PEUBF}
\end{align}
Our task reduces to solving the summation in Eq.~\eqref{Dnf_k}.

To begin with, we consider the simple case of $k=2$, as the case of $k=1$ is solved in Ref.~\cite{zhuang2020entanglement}. For the case of 2-CPF, we will have two kinds of terms in the summation of Eq.~\eqref{Dnf_k}, one with Hamming distance 4 (corresponding to patterns with no target channels overlapping), and one with Hamming distance 2 (corresponding to patterns with one target channel overlapping), so that
\begin{align}
\sum_{{\bs i}\neq {\bs j}, 2-CPF} f^{{\rm hamming}(\bs{i},\bs{j})}=\binom{n}{4} \binom{4}{2} f^4+6\binom{n}{3} f^2.
\end{align}
As further example, consider 2-CPF with $4$ pixels. There are 6 configurations $1100,1010,1001,0110,0101,0011$. In total we have $6\times 5=30$ terms to sum up. There are 6 terms with $f^4$: $(1100,0011),(1010,0101),(1001,0110)$ and their reverse. There are 24 terms with $f^2$, for example $(1100,1010)$.

In general, for k-CPF with $n$ pixels, we have $\binom nk$ configurations of patterns. Therefore, Eq.~\eqref{Dnf_k} is a summation of $\binom nk\left(\binom nk-1\right)$ terms. There are terms with Hamming distance of from 2 to $2k$. By counting the number of terms with an identical hamming distance, we can compute
\begin{align}
&\sum_{{\bs i}\neq {\bs j}, k-CPF} f^{{\rm hamming}(\bs{i},\bs{j})}= \sum_{t=k+1}^{2k} \binom{n}{t} \binom{k}{2k-t}\binom{t}{k} f^{2(t-k)} \nonumber
\\
&=\frac{\left(1+k\right)\binom{n}{k+1}}{n-k}\left({}_2F_1\left(-k,k-n,1,f^2\right)-1\right),
\end{align}
where ${}_2F_1$ is the standard hypergeometric function. 


Therefore we can solve the function $D_n^k$ as
\begin{align}
	D_n^k(f) = 
{}_2F_1\left(-k,k-n,1,f^2\right)-1,
\end{align}
and then evaluate the bounds through Ineqs.~\eqref{PELBF} and~\eqref{PEUBF}.

For large $M$ we may expand $D_n^k$ as 
\begin{equation}
	D_n^k = k(n-k) f^2 + O(f^4)~,		
\end{equation}
and for large $n$ and $k$ we may write $\binom nk \approx 2^{nH(k/n)}$, 
where $H(p) = -p\log_2(p) -(1-p)\log_2(1-p)$ is the binary entropy function.
Employing these approximations in Eqs.~\eqref{PELBF}-\eqref{PEUBF} we find Eq.~\eqref{kCPF}.

\section{Quantum and classical fidelities}\label{a:fid}

In the case of a quantum source, the input state is a two-mode Gaussian state, 
described by the quadrature operators $Q=(x_1,p_1,x_2,q_2)$. 
A TMSV state $\Phi_{N_S}$ has zero first moments and covariance matrix (CM) 
$V_{jk} = \langle\{Q_j,Q_k\}\rangle/2$ given by 
\begin{equation}
	V_{\rm input}=\frac12\left(
\begin{array}
[c]{cc}%
\mu I & \mu^{\prime}Z\\
\mu^{\prime}Z & \mu I
\end{array}
\right)  ,~%
\begin{array}
[c]{l}%
\mu=2N_{S}+1\\
\mu^{\prime}=\sqrt{\mu^{2}-1},%
\end{array}
\end{equation}
where $I$ is the 2x2 identity matrix and $Z$ is the Pauli $Z$ operator. In the above 
expression the variance of the vacuum noise is $1/2$, and $N_S$ is the mean number of thermal photons in each mode.
If we now apply $\mathcal I\otimes\mathcal E_\eta$ to the TMSV state, we 
get a Gaussian output state with CM
\begin{align}
	V(\eta) &  =\frac12\left(
\begin{array}
[c]{cc}%
\mu(1)I & \sqrt{\eta}\mu^{\prime}Z\\
\sqrt{\eta}\mu^{\prime}Z & \mu(\eta)I
\end{array}
\right)  \label{Voutput}
\end{align}
where%
\begin{equation}
\mu(\eta):=\eta\mu+(1-\eta).\label{mueta}%
\end{equation}
The fidelity $F(\rho_W,\rho_B)$ can be computed using the formulae from Refs.~\cite{Fidelity,marian2012uhlmann}, 
and depends on the invariants 
\begin{align}
	\label{asdD}
	\Delta&=\det(V(\eta_B)+V(\eta_W))=\\&=
	\left(N_S^2 \left(\sqrt{\eta_W}-\sqrt{\eta_B}\right)^2-2 N_S \left(\sqrt{\eta_W \eta_B}-1\right)+1\right)^2,
	\nonumber
	\\ 
	\Gamma &= 2^4\det(\Omega V(\eta_B)\Omega V(\eta_W)-I\otimes I/4)
	\label{asdG}
			 \\&=\Delta+ 
			 4 (\eta_W-1) (\eta_B-1) N_S^2 
			 \left(N_S \left(1-\sqrt{\eta_W \eta_B }\right)+1\right)^2,
			 \nonumber
			 \\
	\Lambda &= 2^4 \det (V(\eta_B) + i\Omega/2)\det (V(\eta_W) + i\Omega/2)  =0,\label{asdL}
\end{align}
where $\Omega = (i Y)^{\oplus 2}$ and $Y$ is the Pauli $Y$ operator.
In terms of the above quantities the fidelity can be written as 
\begin{align}
	F(\rho_W,\rho_B) &= \sqrt{\frac1{\sqrt\Gamma+\sqrt\Lambda -\sqrt{(\sqrt\Gamma+\sqrt\Lambda)^2-\Delta}}}
								\nonumber \\ &= 
								\frac1{\sqrt{\frac{\sqrt{\Gamma}+\sqrt{\Lambda}+\sqrt\Delta}2}
								-\sqrt{\frac{\sqrt{\Gamma}+\sqrt\Lambda-\sqrt\Delta}2}},
\end{align}
where in the second line we simplify the first expression, derived in 
\cite{Fidelity,marian2012uhlmann}, by 
explicitly evaluating the square root. 
Inserting the values from Eqs.~\eqref{asdD}-\eqref{asdL}
we get the 
result shown in Eq.~\eqref{fidelityQ}. The tightness of the inequality in Eq.~\eqref{fidelityQ} can be 
proven by setting $N_{\rm tot}=M N_S$ and using $e^x=\lim_{M\to\infty}(1+x/M)^M$
\begin{equation}
	\lim_{M\to\infty} \left(\frac{1}{1+N_{\rm tot}\Delta_Q/M}\right)^M = e^{-N_{\rm tot} \Delta_q}~.
\end{equation}

Now let us compute the output fidelity corresponding to an input coherent state $D(\sqrt{N_s})\ket{0}$, where $D$ is the displacement operator. 
The action of the channel with transmissivity $\eta$ is described by the Heisenberg evolution 
\begin{equation}
	a^\dagger \to \sqrt{\eta}a^\dagger + \sqrt{1-\eta}b^\dagger,
\end{equation}
where $b$ describes the photons in the (vacuum) bath. After the application of the channel we get a new 
coherent state $D(\sqrt{\eta N_S}) \ket 0$, so the fidelity is 
\begin{align}
	\nonumber
	F_{\rm cl} &= |\bra 0 D(\sqrt{\eta_BN_S})^\dagger D(\sqrt{\eta_WN_S})\ket 0| =\\&= 
	\nonumber
	\left|e^{-\frac{N_S\eta_B}2}
	e^{-\frac{N_S\eta_W}2} \sum_n \frac{(N_S\sqrt{\eta_B\eta_W})^n}{n!}\right|
=					\\ &= e^{-\frac{N_s}2(\sqrt{\eta_B}-\sqrt{\eta_W})^2},
\end{align}
which is the result shown in Eq.~\eqref{fidelityC}.

\section{Quantum pattern recognition with pretty-good measurements}\label{s:pgm}
A pretty good solution to the optimization problem \eqref{Equantum} can be 
obtained with pretty good measurements \cite{PGM1,PGM2,PGM3}, which are defined by 
$\Pi_c^{\rm PGM} = \rho_{\rm tot}^{-1/2} P(c) \rho_c^M\rho_{\rm tot}^{-1/2}$,
where $P(c,\bs i)=P(\bs i|c)P(c)$ via the Bayes rule, 
$\rho_c^M = \sum_{\bs i} P(\bs i|c) \rho_{\bs i}^{\otimes M}$ and $\rho_{\rm tot} = \sum_c P(c) \rho_c^M = 
\sum_{c,\bs i} P(c,\bs i) \rho_{\bs i}^{\otimes M}$. 
Using these measurements, we may find an upper bound \cite{Barnum} for $E^{\rm Q}$ and,
by rewriting $E^{Q} = \min_{\{\Pi_c\}} \sum_{c\neq \tilde c} P(c) \Tr[\Pi_{\tilde c}\rho_c^M]$,
we may find a lower bound as in \eqref{BarnumMontanaro}
using similar techniques \cite{Montanaro}. 
 These bounds read
\begin{align}
	E^{\rm Q} &\geq 
	\frac{1}{2}\sum_{c\neq c'} P(c)P(c')
F(\rho_c^M,\rho_{c'}^M)^2,
\\
	E^{\rm Q} &
\leq\sum_{c\neq c'}\sqrt{P(c)P(c')}
F(\rho_c^M,\rho_{c'}^M).
\label{qpUB}
\end{align}
The lower bound can be simplified thanks to the concavity of $F^2$ \cite{jozsa1994fidelity}, so as to get
\begin{align}
	E^{\rm Q} &\geq 
	\frac{1}{2}\sum_{c\neq c'} \sum_{\bs{i},\bs{i'}}P(c,\bs i)P(c',\bs i')
	F(\rho_{\bs i},\rho_{\bs i'})^{2M},
 \label{BarnumMontanaroLearning}
\end{align}
while the same simplification cannot be obtained for the upper bound. 
As a consequence, the upper bound cannot be directly computed, as $\rho_c^M$ is not 
a quantum Gaussian state. 
In order to introduce a computable upper bound we prove the 
following lemma, generalizing methods from~\cite{Barnum,montanaro2019pretty}.

\noindent {\bf Lemma}: Let us fix a function $f$. Then 
using pretty good measurements we can derive the following bound 
for the probability of error 
\begin{align}
	p_{\rm E} &= \sum_{x,y : f(x)\neq f(y)} p_x \Tr[ \Pi_y\rho_x] 
	\label{Pe}
					\\&\leq 
	\sum_{x,y : f(x)\neq f(y)} \sqrt{p_x p_y} F(\rho_x,\rho_y)~.
	\label{Montagen}
\end{align}

\noindent{\it Proof:} Let us write $\sigma_x=p_x \rho_x = \sum_{k} \lambda_{xk} \ket{\psi_{xk}}\!\bra{\psi_{xk}}$,
where $\lambda_{xk}$ and $\ket{\psi_{xk}}$ form the eigen-decomposition of $\sigma_x$. We also set 
$\Pi_y=\sigma^{-1/2}\sigma_y\sigma^{-1/2}$, $\sigma=\sum_x\sigma_x$,
and define the Gram matrix $G^{xy}_{kl} = \sqrt{\lambda_{xk}\lambda_{yl}}
\bra{\psi_{xk}}{\psi_{yl}}\rangle$. Then the following identities hold (see \cite{montanaro2019pretty} for a proof): 
\begin{align}
	\label{Gnorm2}
	\|\sqrt{G}^{xy}\|_2^2 &= p_x \Tr[ \Pi_y\rho_x]~,\\
	\|G^{xy}\|_1 &= \sqrt{p_x p_y}  F(\rho_x,\rho_y)~,
	\label{Gnorm1}
\end{align}
where $\|A\|_2^2 = \Tr[A^\dagger A]$, $\|A\|_1=\Tr\sqrt{A^\dagger A}$ and $G^{xy}$ is a matrix with elements 
$G^{xy}_{kl}$. Moreover, 
	$\|\sqrt{G}^{xy}\|_2^2 \leq \|G^{xy}\|_1 $ 
	from lemma 4 in \cite{Barnum}. Therefore, 
\begin{align}
	p_{\rm E} &= 
	\sum_{x,y : f(x)\neq f(y)} 	\|\sqrt{G}^{xy}\|_2^2  \leq 
%
	\sum_{x,y : f(x)\neq f(y)} 	\|{G}^{xy}\|_1~,
	\nonumber
\end{align}
which produces \eqref{Montagen} via \eqref{Gnorm1}.
\hfill$\square$

\bigskip

Thanks to the above lemma, we may now derive a simpler upper bound than \eqref{qpUB}.
Indeed, we may rewrite \eqref{Equantum} as in \eqref{Pe} by setting $f(c,\bs i) = c$,
calling $\rho_{c,\bs i}:=\rho_{\bs i}$
and $\Pi_{c'}:=\sum_{\bs i'} \Pi_{c',\bs i'}$ and defining multi-indices $x=(c,\bs i)$ 
and $y=(c',\bs i')$. Then from the above lemma we get
\begin{align}
	E^{\rm Q} &\leq 
	\sum_{c\neq c'} \sum_{\bs{i},\bs{i'}}\sqrt{P(c,\bs i)P(c',\bs i')}
	F(\rho_{\bs i},\rho_{\bs i'})^{M}.
	\label{Equb}
\end{align}
Now we show how to get Ineqs.~\eqref{PatternBounds}. 
Indeed, via  \eqref{BarnumMontanaroLearning} and \eqref{FidIdlers} we may write
$E^{\rm Q} \geq B[F(\rho_B,\rho_W)^{2M}]$ where 
\begin{align}
	B[F] &= \frac12\sum_{c\neq c'} P(c)P(c') B_{cc'}[F], \\ 
	B_{cc'}[F] &= \sum_h P_{cc'}(h) F^h, \\
	P_{cc'}(h) &= \sum_{\bs i,\bs i'} P(\bs i|c)P(\bs i'|c') \delta_{h,{\rm hamming}(\bs i,\bs i')}.
\end{align}
Similarly, we may rewrite the upper bound \eqref{Equb} as 
\begin{align}
	E^{\rm Q} &\leq \nonumber
	K \sum_{c\neq c'} \sum_{\bs{i},\bs{i'}}P(c,\bs i)P(c',\bs i')
	F(\rho_{\bs i},\rho_{\bs i'})^{M} = \\ &=K B[F(\rho_B,\rho_W)^M],
	\label{Kerr}
\end{align}
where $K^{-1}$ is the minimum non-zero value of $\sqrt{P(c,\bs i)P(c',\bs i')}$.
Therefore, 
as long as $P_{cc'}(h)$ is zero for $h<h^{\rm min}_{cc'}$ in the limit of large $M$ we get 
\begin{equation}
	C F(\rho_B,\rho_W)^{2M h_{\rm min}} \lessapprox 
E^{\rm Q}\lessapprox 2K C F(\rho_B,\rho_W)^{M h_{\rm min}},
\nonumber
\end{equation}
up to a constant $C$.

We now focus on the bounds in Eq.~\eqref{PatternBounds1}, which follow 
by approximating $P(c,\bs i)$ via the training distribution 
\begin{equation}
	P(c,\bs i) \simeq \frac1T \sum_{k=1}^T \delta_{c,c_k^{\mathcal T}}
		\delta_{\bs i,{\bs i}_k^{\mathcal T}},
		\label{empdistro}
\end{equation}
and noting that, in \eqref{Equb}, we may use 
$\sqrt{\delta_{ij}+\delta_{kl}}\leq \delta_{ij}+\delta_{kl}$. 
Alternatively, we may employ \eqref{Kerr} and note that $K=T$ for the distribution 
\eqref{empdistro}. 

As for the bounds \eqref{PatternBounds1}, 
let us study the function defined in Eq.~\eqref{Bfunc}, which can be rewritten as 
\begin{equation}
	B_{\mathcal T}[F]= \sum_{c\neq c'} \frac{T_cT_{c'}}{2T^2} B^{\mathcal T}_{cc'}[F]
\end{equation}
with  $T_c= \sum_k \delta_{c,c_k^\mathcal T}$, $T=\sum_c T_c$ and 
\begin{align}\label{Bcc}
	B^{\mathcal T}_{cc'}[F] &= \sum_h P^{\mathcal T}_{cc'}(h) F^h,\\\nonumber
	P^{\mathcal T}_{cc'}(h) &= \frac1{T_cT_{c'}}\sum_{k,k'}
\delta_{c,c_k^{\mathcal T}} 
\delta_{c',c_{k'}^{\mathcal T}} 
\delta_{h, {\rm hamming}({\bs i_k^\mathcal T},{\bs i_{k'}^{\mathcal T}})}.
\end{align}
The probability $P^{\mathcal T}_{01}(h)$ is numerically studied in Fig.~\ref{fig:digits}b for 
the MNIST dataset, where we see that $P^{\mathcal T}_{01}(h)$ closely matches a normal distribution.
Therefore, approximating $P^{\mathcal T}_{cc'}(h)$ as a normal distribution with mean $\mu_{cc'}$ and standard 
deviation $\sigma_{cc'}$ we may write
\begin{align}
	B^{\mathcal T}_{cc'}[F] &\simeq \int_{h^0_{cc'}}^{\infty} p_{\mu_{cc'},\sigma_{cc'}}(h) F^h
					 \\&=
					 \frac{1}{2} F^{\mu_{cc'} } e^{\frac{1}{2} \sigma_{cc'} ^2 \log^2(F)} 
					 \left(\text{erf}\left(\frac{w_{cc'}}{\sqrt{2} \sigma_{cc'} }\right)+1\right),\nonumber
\end{align}
where $p_{\mu,\sigma}(h)$ is the probability density function of a normal 
distribution with mean $\mu$ and standard deviation $\sigma$, 
$w_{cc'}:=\sigma_{cc'}^2 \log (F)+\mu_{cc'} -h^0_{cc'}$ and $h^0_{cc'}$ is the 
minimum value of Hamming distance in the empirical distribution. 
When $\mu\gg \sigma$ we find $B_{cc'}^{\mathcal T}\simeq  F^{\mu_{cc'}}$. 
 On the other hand, for $F\to 0$ we find 
\begin{equation}
	B^{\mathcal T}_{cc'}[F] ~~\stackrel{F\to 0}{\simeq}~~
	\frac{\sigma  F^{h^0_{cc'}} e^{-\frac{({h^0_{cc'}}-\mu_{cc'} )^2}{2 \sigma_{cc'}^2}}}{\sqrt{2 \pi } \left(-w_{cc'}\right)}.
\end{equation}
This leads to the asymptotic decay rate $B_{cc'}[F^M] \propto F^{M h^0_{cc'}}$ for large $M$, up to constants 
and logarithmic corrections. 
Using the latter asymptotic decay expression in  Eq.~\eqref{Bcc}, we find that, for large $M$, only the classes with smallest minimum Hamming distance $h^0_{cc'}$ survive. 
Moreover, we assume that the ratio $T_c/T$ does not scale with $M$, e.g.~for digit 
reconstruction we assume that the number of 4s and 9s are basically constant if we increase the number of images.
As such we may write 
\begin{equation}
	\lim_{M\to\infty} \frac{\log B[F^M]}M = h_{\rm min} \log F,
\end{equation}
where $h_{\rm min} = \min_{c\neq c'} h_{cc'}^0$, and from the bounds \eqref{PatternBounds1}
\begin{equation}
	-h_{\rm min} \log F \leq -\lim_{M\to\infty} \frac{\log E^{\rm Q}_{\mathcal T}}M \leq -2h_{\rm min} \log F .
\end{equation}

%

\clearpage

\section*{Supplementary Materials}

\setcounter{equation}{0}
\renewcommand{\theequation}{S\arabic{equation}}

\subsection*{Multiple Chernoff bound}

Here we show that, in the uniform case, 
we may also employ the multiple quantum Chernoff bound 
\cite{QSDReview,MultiChernoff}. Combining this with known inequalities \cite{PirandolaCB}, we 
find 
\begin{equation}
\frac{F^{2M}_{\rm max}}{2^M} \lesssim p_{\rm err} \lesssim  
F_{\rm max}^{\frac {M}3},
\label{ChernofFid}
\end{equation}
where $F_{\rm max} := \max_{\bs{i}\neq\bs{j}} F_{\bs{i}:\bs{j}}$, 
and the approximations are valid in the large-$M$ limit. The above 
inequality becomes exact for $M\to\infty$, while \eqref{BarnumMontanaro} is 
valid for any $M$.

Following \cite{QSDReview,MultiChernoff} for i.i.d.~hypotheses, we may also consider the following quantum Chernoff bound
\begin{equation}
	\frac13 \xi_{CB}\leq
	\lim_{M\to\infty} -\frac1M\log p_{\rm err} \leq \xi_{CB},
	\label{ChernoffIneq}
\end{equation}
where 
\begin{equation}
	\xi_{CB} = \min_{{\bs i}\neq {\bs j}} [-\log(Q(\rho_{\bs i},\rho_{\bs j}))]
	=-\log Q_{\rm max},
\end{equation}
and 
\begin{align}\label{Qdef}
Q(\rho_{0},\rho_{1})  &  :=\inf_{s}\mathrm{Tr}(\rho_{0}^{s}\rho_{1}^{1-s}%
)\leq\mathrm{Tr}\sqrt{\rho_{0}}\sqrt{\rho_{1}}, \\
Q_{\rm max} & = 
\max_{{\bs i}\neq {\bs j}} Q(\rho_{\bs i},\rho_{\bs j}) \leq \max_{{\bs i}\neq {\bs j}} 
\Tr[\sqrt{\rho_{\bs i}}\sqrt{\rho_{\bs j}}],
	\label{Qmax}
\end{align}
which can be computed for any Gaussian state \cite{PirandolaCB}. 
Using known inequalities \cite{fuchs1999cryptographic,PirandolaCB}
\begin{equation}
	1-\sqrt{1-F(\rho_{\bs i},\rho_{\bs j})^2} \leq Q(\rho_{\bs i},\rho_{\bs j}) 
	\leq F(\rho_{\bs i},\rho_{\bs j})~,
\end{equation}
we may write 
\begin{equation}
	1-\sqrt{1-F_{\rm max}^2} \leq Q_{\rm max} \leq F_{\rm max}~,
\end{equation}
where 
\begin{equation}
	F_{\rm max} = \max_{{\bs i}\neq {\bs j}} F(\rho_{\bs i},\rho_{\bs j})~.
\end{equation}
From the above we finally get the following inequalities 
\begin{equation}
	-\log F_{\rm max}  \leq \xi_{CB}\leq 
	-\log(1-\sqrt{1-F_{\rm max}^2})~,
	\label{boundXiCB}
\end{equation}
\begin{align}
	\label{PerrCB}
	-\frac13\log F_{\rm max}  &\leq 
	\lim_{M\to\infty} -\frac1M\log p_{\rm err} \leq \\&\leq
	-\log(1-\sqrt{1-F_{\rm max}^2})
\nonumber	\leq \\&\leq 
	-\log \frac {F_{\rm max}^2}2,
	\nonumber
\end{align}
where in the last inequality we use 
$1-\sqrt{1-x}\geq x/2$ for $0\leq x\leq 1$, which is tight for $x\simeq 0$. 
In the large-$M$ limit we get the inequalities \eqref{ChernofFid}.
Nonetheless, the exact computation of $Q$ via \eqref{Qdef} may provide a tighter inequality 
via \eqref{ChernoffIneq}.

Moreover, from the definition of the states \eqref{RhoIdlers}, and from \eqref{Qmax} we 
get 
\begin{equation}
	Q(\tilde \rho_i,\tilde \rho_j) = Q(\rho_W,\rho_B)^{{\rm hamming}(\bs{i},\bs{j})},
\end{equation}
and accordingly, since $Q\leq 1$, we find 
\begin{equation}
	Q_{\rm max} = Q(\rho_W,\rho_B)^{ \min_{i\neq j}{\rm hamming}(\bs{i},\bs{j})}
	= Q(\rho_W,\rho_B).
\end{equation}
Together with \eqref{ChernoffIneq}, this gives computable upper (and possibly lower) bounds on the 
error rate. 
Similarly, we may write 
\begin{equation}
	F_{\rm max} = F(\rho_W,\rho_B)^{ \min_{i\neq j}{\rm hamming}(\bs{i},\bs{j})}
	= F(\rho_W,\rho_B),
\end{equation}
so we may study the quantum Chernoff bound via \eqref{boundXiCB} and \eqref{PerrCB}.

An interesting comparison is between the asymptotic performance given by {Eq.~\eqref{PerrCB}}, and the explicit results that we may get from 
from Eq.~\eqref{asympbound}. Indeed, for large-$M$ we get
\begin{equation}
	-\log F \leq -\lim_{M\to\infty} \frac{\log p_{\rm err}}{M} \leq -2\log F,
	\label{asympboundlog}
\end{equation}
which is independent on $n$, as it is {Eq.~\eqref{PerrCB}}. 

\subsection*{Photodetection}
We study the performance of barcode decoding and pattern recognition using photodetection. 
Photodetection is a 
local measurement, where each pixel is probed independently. As such, this 
strategy cannot achieve the lower bound in Eq.~\eqref{perrsimple}, but it is 
nonetheless interesting because it can be implemented experimentally with 
current technology. According to Ref.~\cite{ortolano2020experimental}, the pixel error probability for a coherent-state input 
is given by 
\begin{equation}
	p_{\rm err, cl}^{\rm pixel} = \frac 12\left[1-\frac{\gamma(\eta_B)-\gamma(\eta_W)}{\lfloor n_{\rm th}\rfloor!} \right]~,
\end{equation}
where $n_{\rm th} = N_S(\eta_W-\eta_B)/\log(\eta_W/\eta_B)$, while  $\gamma(\eta) := 
\Gamma(\lfloor n_{\rm th}+1\rfloor, N_S\eta)$ and $\Gamma(x,y)$ is the 
incomplete Gamma function. 

For a TMSV with mean photon number $N_S$ we may write $\ket{\rm TMSV} = \sum_n c_{N_s}(n) \ket{n,n}$ 
with $c_{N} = \frac{N_S^n}{(1+N_S)^{n+1}}$. After the first mode of the TMSV state is transmitted through a channel 
with transmissivity $\eta$, the resulting photon number distribution is 
\begin{equation}
	P(n_1,n_2|\eta) = \begin{cases}
		P_0(n_2) B(n_2,n_1|\eta)  & {\rm when~} n_2\geq n_1, \\ 
		0 &{\rm when~} n_2<n_1,
	\end{cases}
	\label{photodist}
\end{equation}
where $B(n_2,n_1|\eta)=\binom{n_2}{n_1} \eta^{n_1}(1-\eta)^{n_2-n_1}$ and $P_0(n_2) = c_{N_S}^2(n_2)$ is the 
initial distribution. The success probability of distinguishing the two channels 
with transmissivity $\eta_B$ and $\eta_W$ is then 
\begin{align}
	\label{photosum}
	p_{\rm succ,TMSV}^{\rm pixel} =& \frac12 \sum_{n_2=0}^\infty 
	\sum_{n_1=0}^{n_2^{\rm th}(n_1)} P(n_1,n_2|\eta_B) +  
															 \\&+ \frac12 \sum_{n_2=0}^\infty 
	\sum_{n_1=n_2^{\rm th}(n_1)}^\infty P(n_1,n_2|\eta_W),
	\nonumber
\end{align}
where 	$n_2^{\rm th}(n_2) = c n_2$ and 
\begin{equation}
c =  \left(\frac{\log(\eta_W/\eta_B)}{\log[(1-\eta_W)(1-\eta_B)]}+1\right)^{-1}~.
\end{equation}
For $M$ probings, one may substitute the initial distribution in Eq.~\eqref{photodist} 
with the Poisson distribution $P_0(n_2) = e^{-\lambda} \lambda^{n_2}/n_2!$ with 
parameter 
$	\lambda \approx M N_S$~\cite{ortolano2020experimental}.

For large average photon number, namely for large $\lambda$, the sum in Eq.~\eqref{photosum} is 
hard to evaluate. Nonetheless, we may approximate the Poisson distribution $P_0(n_2)$ with a 
normal distribution $n_2 \sim \mathcal N(\lambda,\lambda)$, where $\mathcal N(\mu,\sigma^2)$ 
is the normal distribution with mean $\mu$ and variance $\sigma^2$. Similarly, we may approximate 
the binomial distribution in \eqref{photodist} as $n_1 \sim \mathcal N(n_2 \eta, n_2 \eta(1-\eta))$. 
Introducing the quantity 
\begin{equation}
	g(\lambda, \eta) = \int_0^\infty dn_2 \int_0^{c n_2} d n_1 
	f_{\lambda,\lambda}(n_2) 
	f_{n_2 \eta,n_2\eta(1-\eta)}(n_1)~,
\end{equation}
where $f_{\mu,\sigma^2}(x)$ is the probability density function of $\mathcal N(\mu,\sigma^2)$, and approximating the sum with an integral in \eqref{photosum}, we then get 
\begin{equation}
	p_{\rm succ,TMSV}^{\rm pixel} \simeq  \frac12\left[g(MN_S,\eta_B)+1-g(MN_S,\eta_W)\right]~,
\end{equation}
and accordingly 
\begin{equation}
	p_{\rm err,TMSV}^{\rm pixel} \simeq  \frac12 -\frac12\left[g(MN_S,\eta_B)-g(MN_S,\eta_W)\right]~.
\end{equation}

\subsection*{ Nearest neighbor classifier}

Let us review a few important aspects of the nearest neighbor classifier. 
Using the nearest neighbor classifier \eqref{nnrule} with a training set 
composed of $T$ classified images, 
the expected classification 
error for a new image $\bs i$ with true class $c$ is then 
\begin{equation}
	E_{\rm NN}(T) = \ave_{(c,\bs i),\mathcal T}[L_{c,\tilde c_{\rm NN}(\bs i)}]~,
	\label{experrnn}
\end{equation}
where $T$ is the size of the training set. Under mild conditions, it has 
been proven \cite{cover1967nearest} that in the limit $T\to\infty$, 
the expected classification error $E_{\rm NN}=\lim_{T\to\infty}E_{\rm NN}(T)$ 
satisfies 
\begin{equation}
	E_{\rm B}\leq E_{\rm NN} \leq E_{\rm B}\left(2-\frac{K}{K-1}E_{\rm B}\right)
	\leq 2E_{\rm B}~,
	\label{bayesnn}
\end{equation}
where $E_{\rm B}$ is the Bayes rate and $K$ is the number of classes. 
Therefore, the expected error from the nearest 
neighbor classifier is at most twice the Bayes rate, irrespective of the number 
of classes.

To study the error for finite $T$ in \eqref{experrnn} we first note that even the training set 
is made by samples, so the predicted nearest neighbour classifier $\tilde c$ 
is a random variable that depends on the training set $\mathcal T$. (Note that we have removed the subscript NN from $\tilde c$ to simplify the notation.) Therefore, 
we may rewrite Eq.~\eqref{experrnn} as 
\begin{align}
	E_{\rm NN}(T) &= \ave_{(c,\bs i),\tilde c,\mathcal T}\left[L_{c,\tilde c}P(\tilde c|\bs i,\mathcal T)\right]
	\nonumber\\&
	= \ave_{\bs i,\mathcal T}\left[\sum_{c\neq \tilde c} P(c|\bs i)P(\tilde c|\bs i,\mathcal T)\right].
	\label{avefinite}
\end{align}
We know 
from Eq.~\eqref{nnrule} that $\tilde c$ is equal to one element of the training set. Let us 
call $\tilde {\bs i}$ all possible images that have class $\tilde c$. Then by the law 
of total probability 
\begin{equation}
	P(\tilde c|\bs i,\mathcal T)=\sum_{\tilde{\bs i}}P(\tilde c|\tilde{\bs i})
	P_{\rm NN}(\tilde{\bs i}|\bs i,\mathcal T),
	\label{Pnn}
\end{equation}
where $P_{\rm NN}$ is the probability that, given an image $\bs i$, 
in the training set $\mathcal T$ the closest image to ${\bs i}$ is $\tilde {\bs i}$.

In the next sections  we then 
show that $P_{\rm NN}(\tilde{\bs i}|\bs i,\mathcal T)\to\delta_{\bs i,\tilde{\bs i}}$ in the limit 
$T\to\infty$. 
The explanation of this limit is straightforward: since the set of images is finite, in the 
limit of large training sets the probability of finding the image $\bs i$ inside the training set 
approaches 1 and, accordingly, $P_{\rm NN}(\tilde {\bs i}|\bs i)\to \delta_{\bs i,\tilde{\bs i}}$.
Finally, since the number of classes is smaller than the dimension of the set of images, 
we may choose the mapping $P(c|\bs i)$ to be deterministic and unique, e.g. $P(c|\bs i)=\delta_{c,f(\bs i)}$,
where $f$ is a function that assigns a class $c$ to $i$. With these assumptions we find 
\begin{equation}
	\lim_{T\to\infty} E_{\rm NN}(T)= 0~.
\end{equation}
The above result is indeed consistent with \eqref{bayesnn}. 
In fact, for finite dimensional spaces, each image has a unique class associated, so that the Bayes rate is zero.

\subsection*{Nearest neighbor classifier with reading error}
Here we generalize the error \eqref{avefinite} to the case where the input is a reconstructed image ${\bs i'}$, which corresponds to the original image ${\bs i}$ up to an error probability $p_{\rm read}({\bs i}'|\bs i)$ as in \eqref{measupattern}. 
When $p_{\rm read}({\bs i}'|\bs i)=\delta_{{\bs i}',\bs i}$ the reconstruction is perfect,
and the definition \eqref{avefinite} applies. 
On the other hand, when ${\bs i}'\neq {\bs i}$, depending on the noise levels, 
classification algorithms may output the wrong class $\tilde c$. 
Here we focus on the nearest neighbor classifier and generalize
the expected classification error \eqref{avefinite} as 
\begin{align}
	E_{\rm NN}&= \sum_{\bs i'}\ave_{(c,\bs i),\tilde c, \mathcal T}\left[ 
	L_{c,\tilde c} P(\tilde c| {\bs i}',\mathcal T) p_{\rm read}({\bs i}'|\bs i)
	\right]
	\label{avequantum}
					\\&= \nonumber
					1-\sum_c \sum_{\bs i'}\ave_{\bs i, \mathcal T}\left[ 
						P(c|\bs i)P(c| {\bs i}',\mathcal T) p_{\rm read}({\bs i}'|\bs i)
	\right]~.
\end{align}
In the above equation, we assume that $(c,\bs i)$ are sampled from the unknown distribution $P(c,\bs i)$ 
that describes all possible images and their {\it true} label. As in the previous section, we assume 
that the training set is made by samples from $P(c,\bs i)$. Then, Eq.~\eqref{avequantum} can be 
interpreted as follows: we sample a new physical image $\bs i$ and the corresponding true label $c$, which is 
unknown to us. This physical image is measured by a sensor and we get a reconstructed image ${\bs i}'$ 
with probability \eqref{measupattern}. Then we apply the nearest neighbor classifier to find the closest 
image $\tilde{\bs i}$ to $\bs i'$ from the training set, and get its corresponding label $\tilde c$. 
A {\it loss} matrix $L_{c,\tilde c}$ is the used to weigh the error when $\tilde c\neq c$. 

Thanks to the above definition, the optimal measurement in \eqref{measupattern}
is then the one minimizing the expected classification error
\begin{equation}
	E^*_{\rm NN} = \min_{ \{ \Pi_{\bs i'}\}} E_{\rm NN}. 
\end{equation}
Employing \eqref{Pnn} and taking expectation values with respect to the training set 
we find 
\begin{align*}
	E_{\rm NN}&= 1-\sum_c \sum_{\bs i',\tilde{\bs i}}\ave_{\bs i}\left[ 
						P(c|\bs i)P(c|\tilde{\bs i})P_{\rm NN}(\tilde{\bs i}| {\bs i}') p_{\rm read}({\bs i}'|\bs i)
	\right]~.
\end{align*}
If a given image belongs to a single class, as it is the case with the nearest neighbor 
classifier, we may write $P(c|\bs i)=\delta_{c,f(\bs i)}$ for a certain classifier function 
$f$, and accordingly
\begin{equation}
	\sum_c P(c|\bs i)P(c|\bs j)=\delta_{f(\bs i),f(\bs j)}~.
	\label{deltaP}
\end{equation}
so that we may write 
\begin{align}
	E_{\rm NN}&= 1-
	\sum_{\bs i,\bs i',\tilde{\bs i}} \Delta({\bs i,\tilde{\bs i}})
	P_{\rm NN}(\tilde{\bs i}| {\bs i}',\mathcal T) p_{\rm read}({\bs i}'|\bs i) \pi_{\bs i}
	~,
	\label{ennmain}
\end{align} 
For the nearest neighbor classifier, $f$ 
is the function that associates to an image from either the training or testing set 
the corresponding class. For instance, in Eq.~\eqref{ennmain} the image $\bs i$ is 
from the testing set, say with class $c$, and the image $\tilde {\bs i}$ is the nearest 
neighbor to the noisy reconstructed image, and accordingly $\tilde{\bs i}$ is from the 
training set. Calling $\tilde c$ the corresponding class from the training set, 
$\Delta({\bs i},\tilde{\bs i})=\delta_{c,\tilde c}$.
Moreover, in the previous section we have seen that, when $T$ is much bigger
than the dimension of the image space ($T\gg 2^N$ in this case), we may write
$P_{\rm NN}(\bs j|\bs i) = \delta_{\bs i,\bs j}$ 
and we finally get
\begin{equation}
	E_{\rm NN} \stackrel{T\gg 2^N}\approx  
	\ave_{\bs i}\left[ \sum_{\bs i'}(1-\delta_{f({\bs i}),f({\bs i'})})
	p_{\rm read}(\bs i'|\bs i)\right] \leq p_{\rm err}~.
	\label{ennlarged}
\end{equation}
Therefore, in the limit $T\to\infty$ the classification error is smaller than the 
error in the state discrimination \eqref{success}. 

\subsection*{Finite sample-error in the nearest neighbor classifier}

Here we explictly show that $P_{\rm NN}(\tilde{\bs i}|\bs i,\mathcal T)\to\delta_{\bs i,\tilde{\bs i}}$ in the limit 
$T\to\infty$. 
To prove this, we 
study the statistics of $d_{k} = {\rm dist}(\bs i_k^{\mathcal T},\bs i)$ 
and define $d_{\rm min}=\min_k d_k$, where we employ the Hamming distance. 
If the samples $\bs i_k$ are independent and identically distributed we get 
$P_T(d_{\rm min}{\leq}D) \equiv \ave_{\mathcal T} P(d_{\rm min} {\leq} D|\bs i,\mathcal T)$ and 
\begin{align}
	P_T(d_{\rm min} {\leq} D|\bs i,\mathcal T) &= 1-P_T(d_1{>}D,\dots,d_T{>}D|\bs i)
 \nonumber												 \\ &= 1-(1-P(d{\leq} D|\bs i))^T~,
	\label{pminleqd}
\end{align}
where
\begin{equation}
	P(d{\leq} D|\bs i) = \sum_{k=1}^D P({\rm dist}(\bs i',\bs i){=}k|\bs i)
\end{equation}
is the probability that a new image $\bs i'$ has at most distance $D$ from the given image $\bs i$. 
From Eq.~\eqref{pminleqd} we get
\begin{align}
	P_T(d_{\rm min} {=}0|\bs i) &\equiv P_T(d_{\rm min} {\leq}0|\bs i)=
																 \\&\nonumber =
																 (1-(1-P(d{=}0|\bs i))^T)~,
\end{align}
and for any $D>0$ we write
\begin{equation}
	P_T(d_{\rm min} {=}D|\bs i) \equiv P(d_{\rm min} {\leq}D|\bs i)
	-P(d_{\rm min} {\leq}D{-}1|\bs i)~.
\end{equation}
From the above equation and from \eqref{pminleqd}, it is clear that, as long as
$P(d{=}0|\bs i)>0$, for $T\to\infty$ we find
\begin{align}
	P_T(d_{\rm min}{=}0|\bs i) &\to1,&
	P_T(d_{\rm min}{>}0|\bs i) &\to0.
	\label{limit}
\end{align}
In particular, if the set of images is finite and of dimension $N$, then
$P(d{=}0|\bs i)\approx N^{-1}$ and a training set with $T\gg N$ is required 
to approximate the limit~\eqref{limit}.
For any $\bs i$, the solution of $d_{\rm min}=0$ is unique and, therefore, we have that $\bs i_k^{\mathcal T}=\bs i$ for some $k$. 

We may get the above result more explicitly by defining an order between images 
\begin{align}
	\bs j &\preceq_{\bs i} \bs k & \Leftrightarrow  &&
	{\rm dist}(\bs i,\bs j) &\leq {\rm dist}(\bs i,\bs k)~,
\end{align}
so that 
\begin{align}
	P(\bs j \preceq_{\bs i} \bs k|\bs i) &= \sum_{
	{\rm dist}(\bs i,\bs j) \leq {\rm dist}(\bs i,\bs k)}  p_{\bs j}~,
	\\
	P(\bs j \prec_{\bs i} \bs k|\bs i) &= \sum_{
	{\rm dist}(\bs i,\bs j) < {\rm dist}(\bs i,\bs k)}  p_{\bs j}~.
\end{align}
With the above definitions, by calling $P_{\rm NN}(\tilde{\bs i}|\bs i)= 
\ave_{\mathcal T} P_{\rm NN}(\tilde{\bs i}|\bs i,\mathcal T)$, we get
\begin{align}
	P_{\rm NN}(\tilde{\bs i}|\bs i) &:= P_T(\bs j \preceq_{\bs i} \tilde{\bs i}|\bs i)
	- P_T(\bs j \prec_{\bs i} \tilde{\bs i}|\bs i)~,
															 \\ 
	P_T(\bs j \preceq_{\bs i} \tilde{\bs i}|\bs i) & = 1-(1-
	P(\bs j \preceq_{\bs i} \tilde{\bs i}|\bs i))^T~,
\end{align}
and, in particular, for $\tilde{\bs i}=\bs i$ we find  
\begin{align}
	P_{\rm NN}(\bs i|\bs i) := 1-(1-p_{\bs i})^T~.
	\label{pnnT}
\end{align}


\end{document}